\documentclass[10pt,graphicx,subfigure,axodraw,cleverref]{article}
\usepackage{amsfonts}
\usepackage{amssymb}
\usepackage[usenames,dvipsnames]{color}
\usepackage{graphicx}
\usepackage{amsmath}
\usepackage{amsfonts}
\usepackage{amssymb}
\usepackage{cleveref}
\def\rmuu{\gamma^{\mu}}
\def\rmud{\gamma_{\mu}}
\def\PL{{1-\gamma_5\over 2}}
\def\PR{{1+\gamma_5\over 2}}
\def\sinW2{\sin^2\theta_W}
\def\AEM{\alpha_{EM}}
\def\mul{M_{\tilde{u} L}^2}
\def\mur{M_{\tilde{u} R}^2}
\def\mdl{M_{\tilde{d} L}^2}
\def\mdr{M_{\tilde{d} R}^2}
\def\mz2{M_{z}^2}
\def\c2b{\cos 2\beta}
\def\au{A_u}
\def\ad{A_d}
\def\cob{\cot \beta}
\def\v#1{v_#1}
\def\tb{\tan\beta}
\def\epem{$e^+e^-$}
\def\KK{$K^0$-$\overline{K^0}$}
\def\wi{\omega_i}
\def\xj{\chi_j}
\def\Wmu{W_\mu}
\def\Wnu{W_\nu}
\def\m#1{{\tilde m}_#1}
\def\mH{m_H}
\def\mw#1{{\tilde m}_{\omega #1}}
\def\mx#1{{\tilde m}_{\chi^{0}_#1}}
\def\mc#1{{\tilde m}_{\chi^{+}_#1}}
\def\mwi{{\tilde m}_{\omega i}}
\def\mxi{{\tilde m}_{\chi^{0}_i}}
\def\mci{{\tilde m}_{\chi^{+}_i}}

\def\ch{{\tilde\chi^{+}_1}}
\def\c2{{\tilde\chi^{+}_2}}

\def\tt{{\tilde\theta}}

\def\tp{{\tilde\phi}}

\def\mz{M_z}
\def\sw{\sin\theta_W}
\def\cw{\cos\theta_W}
\def\cb{\cos\beta}
\def\sb{\sin\beta}
\def\rwi{r_{\omega i}}
\def\rxj{r_{\chi j}}
\def\rfp{r_f'}
\def\Kik{K_{ik}}
\def\Fq2{F_{2}(q^2)}
\def\f{\({\cal F}\)}
\def\d1{{\f(\tilde c;\tilde s;\tilde W)+ \f(\tilde c;\tilde \mu;\tilde W)}}
\def\tw{\tan\theta_W}
\def\sec2w{sec^2\theta_W}
\begin{document}
%This is dvips(k) 5.86 Cobegin{document}
\baselineskip 18pt
%t
\def\today{\ifcase\month\or
 January\or February\or March\or April\or May\or June\or
 July\or August\or September\or October\or November\or December\fi
 \space\number\day, \number\year}
\def\thebibliography#1{\section*{References\markboth
 {References}{References}}\list
 {[\arabic{enumi}]}{\settowidth\labelwidth{[#1]}
 \leftmargin\labelwidth
 \advance\leftmargin\labelsep
 \usecounter{enumi}}
 \def\newblock{\hskip .11em plus .33em minus .07em}
 \sloppy
 \sfcode`\.=1000\relax}
\let\endthebibliography=\endlist
\def\lsim{\ ^<\llap{$_\sim$}\ }
\def\gsim{\ ^>\llap{$_\sim$}\ }
\def\r2{\sqrt 2}
\def\beq{\begin{equation}}
\def\eeq{\end{equation}}
\def\beqn{\begin{eqnarray}}
\def\eeqn{\end{eqnarray}}
\def\rmuu{\gamma^{\mu}}
\def\rmud{\gamma_{\mu}}
\def\PL{{1-\gamma_5\over 2}}
\def\PR{{1+\gamma_5\over 2}}
\def\sinW2{\sin^2\theta_W}
\def\AEM{\alpha_{EM}}
\def\mul{M_{\tilde{u} L}^2}
\def\mur{M_{\tilde{u} R}^2}
\def\mdl{M_{\tilde{d} L}^2}
\def\mdr{M_{\tilde{d} R}^2}
\def\mz2{M_{z}^2}
\def\c2b{\cos 2\beta}
\def\au{A_u}         
\def\ad{A_d}
\def\cob{\cot \beta}
\def\v#1{v_#1}
\def\tb{\tan\beta}
\def\epem{$e^+e^-$}
\def\KK{$K^0$-$\bar{K^0}$}
\def\wi{\omega_i}
\def\xj{\chi_j}
\def\Wmu{W_\mu}
\def\Wnu{W_\nu}
\def\m#1{{\tilde m}_#1}
\def\mH{m_H}
\def\mw#1{{\tilde m}_{\omega #1}}
\def\mx#1{{\tilde m}_{\chi^{0}_#1}}
\def\mc#1{{\tilde m}_{\chi^{+}_#1}}
\def\mwi{{\tilde m}_{\omega i}}
\def\mxi{{\tilde m}_{\chi^{0}_i}}
\def\mci{{\tilde m}_{\chi^{+}_i}}
\def\mz{M_z}
\def\sw{\sin\theta_W}
\def\cw{\cos\theta_W}
\def\cb{\cos\beta}
\def\sb{\sin\beta}
\def\rwi{r_{\omega i}}
\def\rxj{r_{\chi j}}
\def\rfp{r_f'}
\def\Kik{K_{ik}}
\def\Fq2{F_{2}(q^2)}
\def\f{\({\cal F}\)}
\def\d1{{\f(\tilde c;\tilde s;\tilde W)+ \f(\tilde c;\tilde \mu;\tilde W)}}
%%%%%%%%%%%%%%%%%%%%%%%%%%%%%%%%%%
\def\tw{\tan\theta_W}
\def\sec2w{sec^2\theta_W}
%%%%%%%%%%%%%%%%%%%%%%%%%%%%%%%%%%
\def\ch{{\tilde\chi^{+}_1}}
\def\c2{{\tilde\chi^{+}_2}}

\def\tt{{\tilde\theta}}

\def\tp{{\tilde\phi}}

\def\mz{M_z}
\def\sw{\sin\theta_W}
\def\cw{\cos\theta_W}
\def\cb{\cos\beta}
\def\sb{\sin\beta}
\def\rwi{r_{\omega i}}
\def\rxj{r_{\chi j}}
\def\rfp{r_f'}
\def\Kik{K_{ik}}
\def\Fq2{F_{2}(q^2)}
\def\f{\({\cal F}\)}
\def\d1{{\f(\tilde c;\tilde s;\tilde W)+ \f(\tilde c;\tilde \mu;\tilde W)}}

\def\b{${\cal{B}}(\mu\to {e} \gamma)$~}
\def\bb{${\cal{B}}(\tau\to {\mu} \gamma)$~}

%%%%%%%%%%%%%%%%%%%%%%%%%%%%%%%%%%
\def\tw{\tan\theta_W}
\def\sec2w{sec^2\theta_W}
\newcommand{\pn}[1]{{\color{red}{#1}}}
%%%%%%%%%%%%%%%%%%%%%%%%%%%%%%%%%%

\begin{titlepage}

\begin{center}
{\large {\bf 
$\mu\to e \gamma$ Decay  in an MSSM Extension
}}\\
\vskip 0.5 true cm
\vspace{2cm}
\renewcommand{\thefootnote}
{\fnsymbol{footnote}}
 Tarek Ibrahim$^{a}$\footnote{Email: tibrahim@zewailcity.edu.eg}, Ahmad Itani$^{b}$
  and Pran Nath$^{c}$\footnote{Email: nath@neu.edu}  
\vskip 0.5 true cm
\end{center}

\noindent
{a.University of Science and Technology, Zewail City of Science and Technology,}\\
{ 6th of October City, Giza 12588, Egypt\footnote{Permanent address:  Department of  Physics, Faculty of Science,
University of Alexandria, Alexandria,  Egypt}\\
}
{b. Department of Physics, Faculty of Science, Beirut Arab University, 
Beirut, Lebanon\footnote{Email: a.itanis@bau.edu.lb}} \\
{c. Department of Physics, Northeastern University,
Boston, MA 02115-5000, USA} \\
\vskip 1.0 true cm

\centerline{\bf Abstract}
An analysis is given of the decay $\mu \to e+ \gamma$ in an MSSM extension with 
a vectorlike generation. Here mixing with the  mirrors allows the possibility of this
decay. The analysis is done at   the one loop  level with the exchange of charginos and neutralinos 
and of sleptons and mirror sleptons in the loops. A one loop analysis with W and Z boson exchange and mirror leptons and neutrinos is also considered. The effects of CP violating phases from the new sector on the decay $\mu\to e \gamma$  are  analyzed in detail. 
The constraints arising from  the current upper 
limit on the branching ratio ${\cal{B}}(\mu\to {e} \gamma)$ from the MEG experiment of 
 $2.4\times 10^{-12}$ (at 90\% CL) on the parameter space of SUSY models and on vectorlike models
 are explored.  
Further, the MEG experiment is likely to improve the upper limit by an order of magnitude 
 in the coming years. The improved limits will allow one to probe a much larger domain of the parameter
 space of the extended models.  \\


\medskip

\end{titlepage}

\section{Introduction\label{sec1}}

 Lepton flavor violation provides a new window for physics beyond the standard model.  Since there is no CKM type matrix in the 
 charged leptonic sector, flavor violations involving charged leptons arise via loop corrections  which in particular can produce
 charged lepton flavor violating processes such as $\ell_i^{\pm} \to \ell_j^{\pm} \gamma$.  Recently the MEG experiment ~\cite{Adam:2011ch}
has put the most stringent bound thus far on the lepton flavor violating decay $\mu\to e \gamma$ so that 
 \beqn
    {\cal B}(\mu \to e + \gamma) < 2.4 \times 10^{-12} ~~~~{\rm at ~} 90\% ~{\rm CL} ~~{\rm (MEG)}\,.
     \label{1}
   \eeqn
 In this work we explore the implications of a new leptonic vector generation for the $\mu \to e + \gamma$ decay.  Specifically
 we consider an additional generation of leptons and their mirrors that mix with the three ordinary generations of leptons. Inclusion of a vectorlike generation brings in new sources of CP violation which enter in 
 $\mu\to e \gamma$ decay. These arise from 
  diagrams where  one has charginos and sneutrinos and neutralinos and charged sleptons in the loops.
 Additionally one has diagrams with $W$ and neutrinos and $Z$ and charged leptons in the loops.  Such diagrams can
 produce observable effects and thus the experimental upper limit constrains the parameter space of models. 
 Specifically we will show that the $\mu\to e \gamma$ process can allow one to probe new physics arising from
 the MSSM extension.
 The reason for considering a vectorlike leptonic  generation is the following: 
   First vectorlike generations naturally appear
 in a variety of grand unified models, string models and D brane models and some of these can survive down to low 
 scales\cite{vectorlike}.  Second a vectorlike generation is anomaly free so the good properties of the model as a quantum
 field theory are protected. In previous works ~\cite{Ibrahim:2011im,Ibrahim:2010hv,Ibrahim:2010va,Ibrahim:2008gg,Ibrahim:2009uv,Ibrahim:2012ds,Ibrahim:2014tba,Ibrahim:2014oia,Aboubrahim:2014hya,Aboubrahim:2013yfa,Aboubrahim:2013gfa}
  we have  considered the effects of an extra  vectorlike generation on
 a number of processes and here we extend the analysis to discuss $\mu\to e\gamma$ decay which is one of most
 stringently constrained  lepton flavor violating process. We also investigate the effect of CP phases on the decay $\mu\to e \gamma$.
  Vectorlike multiplets have also been used by other authors (see, e.g., \cite{Babu:2008ge,Liu:2009cc,Martin:2009bg}). 
  Further,  
 $\mu\to e \gamma$ decay has been analyzed in several previous works 
 (see, e.g., \cite{Gabbiani:1988rb,Arnowitt:1990ww,Gabbiani:1996hi,Abada:2008ea,Altmannshofer:2009ne,McKeen:2013dma,Hewett:2012ns}).
   However, none of the previous works  explore the class of models discussed here. \\
    
  The outline of the rest of the paper is as follows: In section 2 we define the model with an extra vectorlike leptonic generation and specify
  the nature of mixings between the extra vectorlike generation and the  three ordinary generations. In section 3
  we give the 
  interactions of the leptons and mirror leptons with the charginos and the neutralinos in the mass diagonal basis. In section 4 we 
  give an analysis of the interactions of the leptons and their mirrors with the W and Z bosons. An analytic analysis of $\mu\to e \gamma$ decay  is given
  in section 5 which includes charginos and neutralinos in the loops as well as W and Z bosons in the loops. A numerical analysis of the 
  $\mu\to e \gamma$ branching ratio is given in section 6. Here it is shown that the vectorlike generation gives a significantly large contribution
  which allows one to probe and constrain the extended model. It is known that CP phases can have a large effect on SUSY loop
  corrections (for a review see~\cite{Ibrahim:2007fb}) and thus  the effect of CP phases on the decay 
  $\mu\to e \gamma$ is also analyzed.
  Conclusions are given in section 7. Details of the scalar mass squared matrices are given in section 8.
  
\section{Extension of MSSM with a Vector Multiplet  \label{sec2}}

In this section we extend MSSM to include a vectorlike  generation which consists of an  ordinary fourth generation of
  leptons, quarks and their mirrors. 
As mentioned in section 1  vectorlike multiplets arise in a variety of unified models some of which could be low lying.
In the analysis below we will assume an extended MSSM with just one  vector multiplet.
 Before proceeding further we
define the notation and give a very brief description of the extended model and  a more detailed
description can be found in the previous works mentioned above. Thus the extended MSSM 
contains a vectorlike multiplet. To fix notation the three generations of leptons are denoted by

{
\begin{align}
\psi_{iL}\equiv   
 \left(\begin{matrix} \nu_{i L}\cr
 ~{l}_{iL}  \end{matrix} \right)  \sim (1,2,- \frac{1}{2}) \ ;  ~~ ~l^c_{iL}\sim (1,1,1)\ ;
 %\nonumber\\
 ~~~ \nu^c_{i L}\sim (1,1,0)\ ;
  ~~~i=1,2,3
\label{2}
\end{align}
}
where the properties  under $SU(3)_C\times SU(2)_L\times U(1)_Y$ are also exhibited.
The last entry in the braces such as $(1,2, -1/2)$ is  
  the value of the hypercharge
 $Y$ defined so that $Q=T_3+ Y$.  These leptons have $V-A$ interactions.
We can now add a vectorlike multiplet where we have a fourth family of leptons with $V-A$ interactions
whose transformations can be gotten from Eq.(\ref{2}) by letting {$i$ run from 1 to 4.}
A vectorlike lepton multiplet also has  mirrors and so we consider these mirror
leptons which have $V+A$ interactions.  {The quantum numbers of the mirrors} are given by

{
\begin{align}
\chi^c\equiv
 \left(\begin{matrix} E_{ L}^c \cr
 N_L^c\end{matrix}\right)  \sim (1,2,\frac{1}{2})\ ; 
~~  E_L \sim  (1,1,-1)\ ;  ~~   N_L \sim (1,1,0).
\label{3}
\end{align}
}

Interesting new physics arises when we allow mixings of the vectorlike generation with
the three ordinary generations.  Here we focus on the mixing of the mirrors in the vectorlike 
generation with the three generations.
Thus the  superpotential of the model allowing for the mixings
among the three ordinary generations and the vectorlike generation is given by

\begin{align}
W&= -\mu \epsilon_{ij} \hat H_1^i \hat H_2^j+\epsilon_{ij}  [f_{1}  \hat H_1^{i} \hat \psi_L ^{j}\hat \tau^c_L
 +f_{1}'  \hat H_2^{j} \hat \psi_L ^{i} \hat \nu^c_{\tau L}
+f_{2}  \hat H_1^{i} \hat \chi^c{^{j}}\hat N_{L}
 +f_{2}'  H_2^{j} \hat \chi^c{^{i}} \hat E_{ L} \nonumber \\
&+ h_{1}  H_1^{i} \hat\psi_{\mu L} ^{j}\hat\mu^c_L
 +h_{1}'  H_2^{j} \hat\psi_{\mu L} ^{i} \hat\nu^c_{\mu L}
+ h_{2}  H_1^{i} \hat\psi_{e L} ^{j}\hat e^c_L
 +h_{2}'  H_2^{j} \hat\psi_{e L} ^{i} \hat\nu^c_{e L}] \nonumber \\
&+ f_{3} \epsilon_{ij}  \hat\chi^c{^{i}}\hat\psi_L^{j}
 + f_{3}' \epsilon_{ij}  \hat\chi^c{^{i}}\hat\psi_{\mu L}^{j}
 + f_{4} \hat\tau^c_L \hat E_{ L}  +  f_{5} \hat\nu^c_{\tau L} \hat N_{L}
 + f_{4}' \hat\mu^c_L \hat E_{ L}  +  f_{5}' \hat\nu^c_{\mu L} \hat N_{L} \nonumber \\
&+ f_{3}'' \epsilon_{ij}  \hat\chi^c{^{i}}\hat\psi_{e L}^{j}
 + f_{4}'' \hat e^c_L \hat E_{ L}  +  f_{5}'' \hat\nu^c_{e L} \hat N_{L}\ ,
 \label{5}
\end{align}
where  $\hat ~$ implies superfields,  $\hat\psi_L$ stands for $\hat\psi_{3L}$, $\hat\psi_{\mu L}$ stands for $\hat\psi_{2L}$
and  $\hat\psi_{e L}$ stands for $\hat\psi_{1L}$.
In \cref{5} we have suppressed terms such as $e^c_4E_L$ etc for simplicity. Their inclusion will
not change our analysis  substantially.
The mass terms for the neutrinos, mirror neutrinos,  leptons and  mirror leptons arise from the term
\beq
{\cal{L}}=-\frac{1}{2}\frac{\partial ^2 W}{\partial{A_i}\partial{A_j}}\psi_ i \psi_ j+\text{H.c.}\,,
\label{6}
\eeq
where $\psi$ and $A$ stand for generic two-component fermion and scalar fields.
After spontaneous breaking of the electroweak symmetry, ($\langle H_1^1 \rangle=v_1/\sqrt{2} $ and $\langle H_2^2\rangle=v_2/\sqrt{2}$),
we have the following set of mass terms written in the 4-component spinor notation
so that
\beq
-{\cal L}_m= \bar\xi_R^T (M_f) \xi_L +\bar\eta_R^T(M_{\ell}) \eta_L +\text{H.c.},
\eeq
where the basis vectors in which the mass matrix is written is given by
\begin{gather}
\bar\xi_R^T= \left(\begin{matrix}\bar \nu_{\tau R} & \bar N_R & \bar \nu_{\mu R}
&\bar \nu_{e R} \end{matrix}\right),\nonumber\\
\xi_L^T= \left(\begin{matrix} \nu_{\tau L} &  N_L &  \nu_{\mu L}
& \nu_{e L} \end{matrix}\right) \ ,\nonumber\\
\bar\eta_R^T= \left(\begin{matrix}\bar{\tau_ R} & \bar E_R & \bar{\mu_ R}
&\bar{e_ R} \end{matrix}\right),\nonumber\\
\eta_L^T= \left(\begin{matrix} {\tau_ L} &  E_L &  {\mu_ L}
& {e_ L} \end{matrix}\right) \ ,
\end{gather}
and the mass matrix $M_f$ is given by

\beqn
M_f=
 \left(\begin{matrix} f'_1 v_2/\sqrt{2} & f_5 & 0 & 0 \cr
 -f_3 & f_2 v_1/\sqrt{2} & -f_3' & -f_3'' \cr
0&f_5'&h_1' v_2/\sqrt{2} & 0 \cr
0 & f_5'' & 0 & h_2' v_2/\sqrt{2}\end{matrix} \right)\ .
\label{7aa}
\eeqn
We define the matrix element $(22)$ of the mass matrix as $m_N$ so that 
\beqn
m_N= f_2 v_1/\sqrt 2.
\eeqn 
The mass matrix is not hermitian and thus one needs bi-unitary transformations to diagonalize it.
We define the bi-unitary transformation so that

\beq
D^{\nu \dagger}_R (M_f) D^\nu_L=\text{diag}(m_{\psi_1},m_{\psi_2},m_{\psi_3}, m_{\psi_4} ).
\label{7a}
\eeq
Under the bi-unitary transformations the basis vectors transform so that
\beqn
 \left(\begin{matrix} \nu_{\tau_R}\cr
 N_{ R} \cr
\nu_{\mu_R} \cr
\nu_{e_R} \end{matrix}\right)=D^{\nu}_R \left(\begin{matrix} \psi_{1_R}\cr
 \psi_{2_R}  \cr
\psi_{3_R} \cr
\psi_{4_R}\end{matrix}\right), \  \
\left(\begin{matrix} \nu_{\tau_L}\cr
 N_{ L} \cr
\nu_{\mu_L} \cr
\nu_{e_L}\end{matrix} \right)=D^{\nu}_L \left(\begin{matrix} \psi_{1_L}\cr
 \psi_{2_L} \cr
\psi_{3_L} \cr
\psi_{4_L}\end{matrix}\right) \ .
\label{8}
\eeqn
%{
In 
$\psi_1, \psi_2, \psi_3, \psi_4$ are the mass eigenstates for the neutrinos,
where in the limit of no mixing
we identify $\psi_1$ as the light tau neutrino, $\psi_2$ as the
heavier mass eigenstate,  $\psi_3$ as the muon neutrino and $\psi_4$ as the electron neutrino.
A similar analysis goes to the lepton mass matrix $M_\ell$ where
\beqn
M_\ell=
 \left(\begin{matrix} f_1 v_1/\sqrt{2} & f_4 & 0 & 0 \cr
 f_3 & f'_2 v_2/\sqrt{2} & f_3' & f_3'' \cr
0&f_4'&h_1 v_1/\sqrt{2} & 0 \cr
0 & f_4'' & 0 & h_2 v_1/\sqrt{2}\end{matrix} \right)\ .
\label{7bb}
\eeqn
In general $f_3, f_4, f_5, f_3', f_4',f_5',  f_3'', f_4'',f_5''$ can be complex and we define their phases
so that

\beqn
f_k= |f_k| e^{i\chi_k}, ~~f_k'= |f_k'| e^{i\chi_k'}, ~~~f_k''= |f_k''| e^{i\chi_k''}\ ;  k=3,4,5\ . 
\eeqn

We introduce now the mass parameter $m_E$ defined by the (22) element of the mass matrix above so that
\beqn
m_E=  f_2' v_2/\sqrt 2.
\eeqn
  Next we  consider  the mixing of the charged sleptons and the charged mirror sleptons.
The mass squared  matrix of the slepton - mirror slepton comes from three sources:  the F term, the
D term of the potential and the soft {SUSY} breaking terms.
Using the  superpotential of  Eq. (4), the mass terms arising from it
after the breaking of  the electroweak symmetry are given by
the Lagrangian
\beq
{\cal L}= {\cal L}_F +{\cal L}_D + {\cal L}_{\rm soft}\ ,
\eeq
where   $ {\cal L}_F$ is deduced from $F_i =\partial W/\partial A_i$, and $- {\cal L}_F=V_F=F_i F^{*}_i$  is given in the appendix, while the ${\cal L}_D$ is given by
\begin{align}
-{\cal L}_D&=\frac{1}{2} m^2_Z \cos^2\theta_W \cos 2\beta \{\tilde \nu_{\tau L} \tilde \nu^*_{\tau L} -\tilde \tau_L \tilde \tau^*_L
+\tilde \nu_{\mu L} \tilde \nu^*_{\mu L} -\tilde \mu_L \tilde \mu^*_L
+\tilde \nu_{e L} \tilde \nu^*_{e L} -\tilde e_L \tilde e^*_L \nonumber \\
&+\tilde E_R \tilde E^*_R -\tilde N_R \tilde N^*_R\}
+\frac{1}{2} m^2_Z \sin^2\theta_W \cos 2\beta \{\tilde \nu_{\tau L} \tilde \nu^*_{\tau L}
 +\tilde \tau_L \tilde \tau^*_L
+\tilde \nu_{\mu L} \tilde \nu^*_{\mu L} +\tilde \mu_L \tilde \mu^*_L \nonumber \\
&+\tilde \nu_{e L} \tilde \nu^*_{e L} +\tilde e_L \tilde e^*_L
-\tilde E_R \tilde E^*_R -\tilde N_R \tilde N^*_R +2 \tilde E_L \tilde E^*_L -2 \tilde \tau_R \tilde \tau^*_R
-2 \tilde \mu_R \tilde \mu^*_R -2 \tilde e_R \tilde e^*_R
\}.
\label{12}
\end{align}
For ${\cal L}_{\rm soft}$ we assume the following form
\begin{align}
-{\cal L}_{\text{soft}}&= M^2_{\tilde \tau L} \tilde \psi^{i*}_{\tau L} \tilde \psi^i_{\tau L}
+ M^2_{\tilde \chi} \tilde \chi^{ci*} \tilde \chi^{ci}
+ M^2_{\tilde \mu L} \tilde \psi^{i*}_{\mu L} \tilde \psi^i_{\mu L}
+M^2_{\tilde e L} \tilde \psi^{i*}_{e L} \tilde \psi^i_{e L}
+ M^2_{\tilde \nu_\tau} \tilde \nu^{c*}_{\tau L} \tilde \nu^c_{\tau L}
 + M^2_{\tilde \nu_\mu} \tilde \nu^{c*}_{\mu L} \tilde \nu^c_{\mu L} \nonumber \\
&+ M^2_{\tilde \nu_e} \tilde \nu^{c*}_{e L} \tilde \nu^c_{e L}
+ M^2_{\tilde \tau} \tilde \tau^{c*}_L \tilde \tau^c_L
+ M^2_{\tilde \mu} \tilde \mu^{c*}_L \tilde \mu^c_L
+ M^2_{\tilde e} \tilde e^{c*}_L \tilde e^c_L
+ M^2_{\tilde E} \tilde E^*_L \tilde E_L
 +  M^2_{\tilde N} \tilde N^*_L \tilde N_L \nonumber \\
&+\epsilon_{ij} \{f_1 A_{\tau} H^i_1 \tilde \psi^j_{\tau L} \tilde \tau^c_L
-f'_1 A_{\nu_\tau} H^i_2 \tilde \psi ^j_{\tau L} \tilde \nu^c_{\tau L}
+h_1 A_{\mu} H^i_1 \tilde \psi^j_{\mu L} \tilde \mu^c_L
-h'_1 A_{\nu_\mu} H^i_2 \tilde \psi ^j_{\mu L} \tilde \nu^c_{\mu L} \nonumber \\
&+h_2 A_{e} H^i_1 \tilde \psi^j_{e L} \tilde e^c_L
-h'_2 A_{\nu_e} H^i_2 \tilde \psi ^j_{e L} \tilde \nu^c_{e L}
+f_2 A_N H^i_1 \tilde \chi^{cj} \tilde N_L
-f'_2 A_E H^i_2 \tilde \chi^{cj} \tilde E_L +\text{H.c.}\}\ .
\label{13}
\end{align}
Here $M_{\tilde e L}, M_{\tilde \nu_e}$ etc are the soft masses and $A_e, A_{\nu_e}$ etc are the trilinear couplings.
The trilinear couplings are complex  and we define their phases so that 
\begin{gather}
A_e= |A_e| e^{i \alpha_{A_e}} \  ,
 ~~A_{\nu_e}=  |A_{\nu_e}|
 e^{i\alpha_{A_{\nu_e}}} \ , 
  \cdots \ .
\end{gather}
From these terms we construct the scalar mass square matrices. These are exhibited in section 8.

\section{Interactions with charginos and neutralinos \label{sec3}}
 In this section we discuss the  interactions in the mass diagonal basis involving charged leptons,
 sneutrinos and charginos.  Thus we have
\begin{align}
-{\cal L}_{\tau-\tilde{\nu}-\chi^{-}} &= \sum_{i=1}^{2}\sum_{j=1}^{8}\bar{\tau}_{\alpha}(C_{\alpha ij}^{L}P_{L}+C_{\alpha ij}^{R}P_{R})\tilde{\chi}^{ci}\tilde{\nu}_{j}+\text{H.c.},
\end{align}
such that

\begin{align}
C_{\alpha ij}^{L}&=g(-\kappa_{\tau}U^{*}_{i2}D^{\tau*}_{R1\alpha} \tilde{D}^{\nu}_{1j} -\kappa_{\mu}U^{*}_{i2}D^{\tau*}_{R3\alpha}\tilde{D}^{\nu}_{5j}-
\kappa_{e}U^{*}_{i2}D^{\tau*}_{R4\alpha}\tilde{D}^{\nu}_{7j}\nonumber\\
&~~~+U^{*}_{i1}D^{\tau*}_{R2\alpha}\tilde{D}^{\nu}_{4j}-
\kappa_{N}U^{*}_{i2}D^{\tau*}_{R2\alpha}\tilde{D}^{\nu}_{2j})
\end{align}

\begin{align}
C_{\alpha ij}^{R}=&g(-\kappa_{\nu_{\tau}}V_{i2}D^{\tau*}_{L1\alpha}\tilde{D}^{\nu}_{3j}-\kappa_{\nu_{\mu}}V_{i2}D^{\tau*}_{L3\alpha}\tilde{D}^{\nu}_{6j}-
\kappa_{\nu_{e}}V_{i2}D^{\tau*}_{L4\alpha}\tilde{D}^{\nu}_{8j}+V_{i1}D^{\tau*}_{L1\alpha}\tilde{D}^{\nu}_{1j}\nonumber\\
&+V_{i1}D^{\tau*}_{L3\alpha}\tilde{D}^{\nu}_{5j}
+V_{i1}D^{\tau*}_{L4\alpha}\tilde{D}^{\nu}_{7j}-\kappa_{E}V_{i2}D^{\tau*}_{L2\alpha}\tilde{D}^{\nu}_{4j}),
\end{align}
with

\begin{align}
(\kappa_{N},\kappa_{\tau},\kappa_{\mu},\kappa_{e})&=\frac{(m_{N},m_{\tau},m_{\mu},m_{e})}{\sqrt{2}m_{W}\cos\beta} , \\~\nonumber\\
%\nonumber\\
(\kappa_{E},\kappa_{\nu_{\tau}},\kappa_{\nu_{\mu}},\kappa_{\nu_{e}})&=\frac{(m_{E},m_{\nu_{\tau}},m_{\nu_{\mu}},m_{\nu_{e}})}{\sqrt{2}m_{W}\sin\beta} .
\end{align}
and
\begin{equation}
U^* M_C V= {\rm diag} (m_{\tilde \chi_1^-}, m_{\tilde \chi_2^-}).
\label{2.4}
\end{equation}

  Next we   discuss the  interactions in the mass diagonal basis involving charged leptons,
 sleptons and neutralinos.  Thus we have

\begin{align}
-{\cal L}_{\tau-\tilde{\tau}-\chi^{0}} &= \sum_{i=1}^{4}\sum_{j=1}^{8}\bar{\tau}_{\alpha}(C_{\alpha ij}^{'L}P_{L}+C_{\alpha ij}^{'R}P_{R})\tilde{\chi}^{0}_{i}\tilde{\tau}_{j}+\text{H.c.},
\end{align}
such that
\begin{align}
%\begin{split}
C_{\alpha ij}^{'L}=&\sqrt{2}(\alpha_{\tau i}D^{\tau *}_{R1\alpha}\tilde{D}^{\tau}_{1j}-\delta_{E i}D^{\tau *}_{R2\alpha}\tilde{D}^{\tau}_{2j}-
\gamma_{\tau i}D^{\tau *}_{R1\alpha}\tilde{D}^{\tau}_{3j}+\beta_{E i}D^{\tau *}_{R2\alpha}\tilde{D}^{\tau}_{4j}
+\alpha_{\mu i}D^{\tau *}_{R3\alpha}\tilde{D}^{\tau}_{5j}\nonumber\\
&-\gamma_{\mu i}D^{\tau *}_{R3\alpha}\tilde{D}^{\tau}_{6j} 
+\alpha_{e i}D^{\tau *}_{R4\alpha}\tilde{D}^{\tau}_{7j}-\gamma_{e i}D^{\tau *}_{R4\alpha}\tilde{D}^{\tau}_{8j})
\end{align}
%\end{split} \\  ~\nonumber
%\begin{split}
\begin{align}
C_{\alpha ij}^{'R}=&\sqrt{2}(\beta_{\tau i}D^{\tau *}_{L1\alpha}\tilde{D}^{\tau}_{1j}-\gamma_{E i}D^{\tau *}_{L2\alpha}\tilde{D}^{\tau}_{2j}-
\delta_{\tau i}D^{\tau *}_{L1\alpha}\tilde{D}^{\tau}_{3j}+\alpha_{E i}D^{\tau *}_{L2\alpha}\tilde{D}^{\tau}_{4j}
+\beta_{\mu i}D^{\tau *}_{L3\alpha}\tilde{D}^{\tau}_{5j}\nonumber\\
&-\delta_{\mu i}D^{\tau *}_{L3\alpha}\tilde{D}^{\tau}_{6j}   
+\beta_{e i}D^{\tau *}_{L4\alpha}\tilde{D}^{\tau}_{7j}-\delta_{e i}D^{\tau *}_{L4\alpha}\tilde{D}^{\tau}_{8j}),
%\end{split}
\end{align}
where

\begin{align}
\alpha_{E i}&=\frac{gm_{E}X^{*}_{4i}}{2m_{W}\sin\beta} \,,  && \beta_{E i}=eX'_{1i}+\frac{g}{\cos\theta_{W}}X'_{2i}\left(\frac{1}{2}-\sin^{2}\theta_{W}\right) \,,\\
\gamma_{E i}&=eX^{'*}_{1i}-\frac{g\sin^{2}\theta_{W}}{\cos\theta_{W}}X^{'*}_{2i} \,,   && \delta_{E i}=-\frac{gm_{E}X_{4i}}{2m_{W}\sin\beta}\,,
\end{align}
and
\begin{align}
\alpha_{\tau i}&=\frac{gm_{\tau}X_{3i}}{2m_{W}\cos\beta} \,,  && \alpha_{\mu i}=\frac{gm_{\mu}X_{3i}}{2m_{W}\cos\beta} \,,  && \alpha_{e i}=\frac{gm_{e}X_{3i}}{2m_{W}\cos\beta}  \,,\\
\delta_{\tau i}&=-\frac{gm_{\tau}X^{*}_{3i}}{2m_{W}\cos\beta} \,, && \delta_{\mu i}=-\frac{gm_{\mu}X^{*}_{3i}}{2m_{W}\cos\beta} \,, && \delta_{e i}=-\frac{gm_{e}X^{*}_{3i}}{2m_{W}\cos\beta}\,,
\end{align}
and where 

\begin{align}
\beta_{\tau i}=\beta_{\mu i}=\beta_{e i}&=-eX^{'*}_{1i}+\frac{g}{\cos\theta_{W}}X^{'*}_{2i}\left(-\frac{1}{2}+\sin^{2}\theta_{W}\right) \,, \\
\gamma_{\tau i}=\gamma_{\mu i}=\gamma_{e i}&=-eX'_{1i}+\frac{g\sin^{2}\theta_{W}}{\cos\theta_{W}}X'_{2i}\,.
\end{align}
Here $X'$ are defined by

\begin{align}
X'_{1i}&=X_{1i}\cos\theta_{W}+X_{2i}\sin\theta_{W}  \,,\\
X'_{2i}&=-X_ {1i}\sin\theta_{W}+X_{2i}\cos\theta_{W}\,,
\end{align}
where $X$ diagonalizes the neutralino mass matrix and is defined by   the relation

\begin{equation}
X^T M_{\tilde \chi^0} X= 
{diag}( m_{\chi_1^0},
 m_{\chi_2^0}, 
 m_{\chi_3^0}, 
 m_{\chi_4^0}) \ .
\end{equation}

\section{Interaction of leptons and mirrors with W and Z bosons\label{sec4}}
In addition to the computation of the supersymmetric loop diagrams, we compute the contributions
arising from the exchange of the W and Z bosons and the leptons and the mirror leptons in the
loops. The relevant interactions needed are given below. For the W boson exchange the
interactions that enter are given by

\begin{align}
-{\cal L}_{\tau W\psi} &= W^{\dagger}_{\rho}\sum_{i=1}^{4}\sum_{\alpha=1}^{4}\bar{\psi}_{i}\gamma^{\rho}[C_{L_{i\alpha}}^W P_L + C_{R_{i\alpha}}^W P_R]\tau_{\alpha}+\text{H.c.}\,,
\end{align}
where

\begin{align}
C_{L_{i\alpha}}^W&= \frac{g}{\sqrt{2}} [D^{\nu*}_{L1i}D^{\tau}_{L1\alpha}+
D^{\nu*}_{L3i}D^{\tau}_{L3\alpha}+D^{\nu*}_{L4i}D^{\tau}_{L4\alpha}] \,, \\
C_{R_{i\alpha}}^W&= \frac{g}{\sqrt{2}}[D^{\nu*}_{R2i}D^{\tau}_{R2\alpha}]\,.
\end{align}
For the Z boson exchange the interactions that enter are given by

\beqn
-{\cal L}_{\tau\tau Z} &= Z_{\rho}\sum_{\alpha=1}^{4}\sum_{\beta=1}^{4}\bar{\tau}_{\alpha}\gamma^{\rho}[C_{L_{\alpha \beta}}^Z P_L + C_{R_{\alpha \beta}}^Z P_R]\tau_{\beta}\,,~~~~~~~~~~~~~~~~~~~~~~~~~~~~~~
\eeqn
 where
\beqn
C_{L_{\alpha \beta}}^Z=\frac{g}{\cos\theta_{W}} [x(D_{L\alpha 1}^{\tau\dag}D_{L1\beta}^{\tau}+D_{L\alpha 2}^{\tau\dag}D_{L2\beta}^{\tau}+D_{L\alpha 3}^{\tau\dag}D_{L3\beta}^{\tau}+D_{L\alpha 4}^{\tau\dag}D_{L4\beta}^{\tau})\nonumber\\
-\frac{1}{2}(D_{L\alpha 1}^{\tau\dag}D_{L1\beta}^{\tau}+D_{L\alpha 3}^{\tau\dag}D_{L3\beta}^{\tau}+D_{L\alpha 4}^{\tau\dag}D_{L4\beta}^{\tau})]\,,
\eeqn
and
\beqn
C_{R_{\alpha \beta}}^Z=\frac{g}{\cos\theta_{W}} [x(D_{R\alpha 1}^{\tau\dag}D_{R1\beta}^{\tau}+D_{R\alpha 2}^{\tau\dag}D_{R2\beta}^{\tau}+D_{R\alpha 3}^{\tau\dag}D_{R3\beta}^{\tau}+D_{R\alpha 4}^{\tau\dag}D_{R4\beta}^{\tau})\nonumber\\
-\frac{1}{2}(D_{R\alpha 2}^{\tau\dag}
D_{R 2\beta }^{\tau}
 )]\,,
\eeqn
where $x=\sin^{2}\theta_{W}$.

\section{The analysis of $\mu \rightarrow e + \gamma$  Branching Ratio \label{sec5} }

\begin{figure}
\begin{center}
\includegraphics[scale=.35]{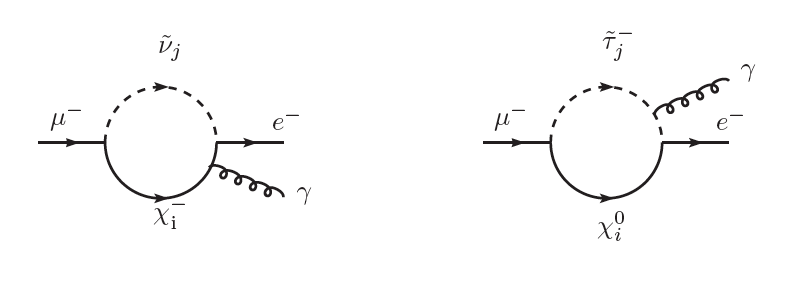}
\includegraphics[scale=.35]{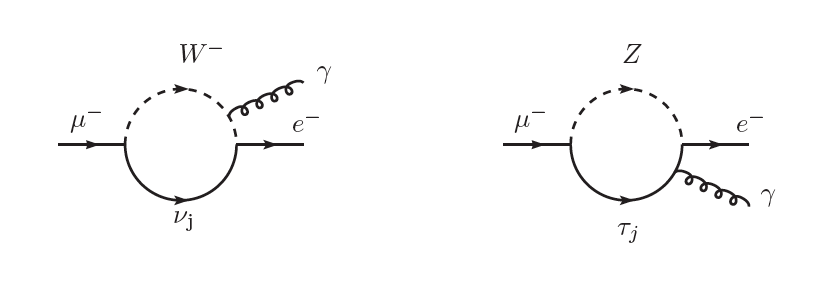}
\caption{\label{muegamma} 
The diagrams that allow the decay of $\mu\to e+\gamma$ via supersymmetric loops involving the chargino  
(top left) and the neutralino (top right) and via $W$ loop (bottom left) and $Z$ loop (bottom right)
 with emission of the photon from the charged particle inside the loop.}
\end{center}
\label{fig1}
\end{figure}

The decay $\mu \rightarrow e + \gamma$
is induced by one-loop electric and 
magnetic  transition dipole moments,  which arise
from the diagrams of Fig.\ref{fig1}.
For an incoming muon of momentum $p$ and a resulting electron of momentum $p'$, we define the amplitude
\beq
<e (p') | J_{\alpha} | \mu (p)> = \bar{u}_{e} (p') \Gamma_{\alpha} u_{\mu} (p) \,,
\label{26}
\eeq
where
\beq
\Gamma_{\alpha} (q) =\frac{F^{\mu e}_2 (q) i \sigma_{\alpha \beta} q^{\beta}}{m_{\mu} +m_{e}}
+\frac{F^{\mu e}_3 (q)  \sigma_{\alpha \beta} \gamma_5 q^{\beta}}{m_{\mu} +m_{e}}+.....
\label{27}
\eeq
with $q = p' -p$ and where $m_f$ denotes the mass of the fermion $f$.
The branching ratio of $\mu \rightarrow e + \gamma$ is given by
\beqn
   {\cal B} (\mu \rightarrow e + \gamma) =\frac{24 \pi^2}{ G^2_F m^2_{\mu} (m_{\mu}+m_{e})^2} \{|F^{\mu e}_2 (0)|^2
+|F^{\mu e}_3 (0)|^2 \}\,,
\label{28}
\eeqn
where the form factors $F^{\mu e}_2$ and $F^{\mu e}_3$ arise from the chargino,
neutralino and vector bosons contributions as follows
\beqn
\label{29}
F^{\mu e}_2 (0) = F^{\mu e}_{2 \chi^+} + F^{\mu e}_{2 \chi^0} + F^{\mu e}_{2 W}+ F^{\mu e}_{2 Z}\,,\\
F^{\mu e}_3 (0) = F^{\mu e}_{3 \chi^+} + F^{\mu e}_{3 \chi^0} + F^{\mu e}_{3 W}+ F^{\mu e}_{3 Z} \,.
\label{29a}
\eeqn

It is also useful to define ${\cal B}_m$ and ${\cal B}_e$ as follows
\begin{align}
\label{bm}
{\cal B }_m (\mu \rightarrow e + \gamma) =\frac{24 \pi^2}{ G^2_F m^2_{\mu} (m_{\mu}+m_{e})^2} |F^{\mu e}_2 (0)|^2\,,\\
{\cal B}_e (\mu \rightarrow e + \gamma) =\frac{24 \pi^2}{ G^2_F m^2_{\mu} (m_{\mu}+m_{e})^2} |F^{\mu e}_3 (0)|^2\,,
\label{be}
\end{align}
where ${\cal B}_m$ is the branching ratio from the  magnetic dipole operator  and ${\cal B}_e$ is the 
 branching ratio from the electric dipole operator. We discuss now the individual contributions to  $F^{\mu e}_{2}$
$F^{\mu e}_{3}$ supersymmetric and non-supersymmetric loops. \\

The chargino contribution  $F^{\mu e}_{2 \chi^+}$ is given by
\beqn
F^{\mu e}_{2 \chi^+}=\sum_{i=1}^2 \sum_{j=1}^8 \bigg[ \frac{m_{\mu}(m_{\mu} +m_{e})}{64 \pi^2 m^2_{\tilde{\chi_i}^+}}\{C^L_{4ij} C^{L*}_{3ij} + C^R_{4ij} C^{R*}_{3ij} \} F_1 \left(\frac{M^2_{\tilde{\nu_j}}}{m^2_{\tilde{\chi_i}^+}}\right)\nonumber\\
+ \frac{(m_{\mu} +m_{e})}{64 \pi^2 m_{\tilde{\chi_i}^+}}
\{C^L_{4ij} C^{R*}_{3ij} + C^R_{4ij} C^{L*}_{3ij} \} F_2\left(\frac{M^2_{\tilde{\nu_j}}}{m^2_{\tilde{\chi_i}^+}}\right)\bigg] \,,
\label{30}
\eeqn
where
\beq
F_1(x)= \frac{1}{3(x-1)^4} \{-2 x^3 -3x^2 +6x -1 +6x^2 \ln x \}
\label{31}
\eeq
and 
\beq
F_2(x)= \frac{1}{(x-1)^3} \{3x^2 -4x +1 -2x^2 \ln x \}\,.
\label{32}
\eeq

\noindent
The neutralino contribution  $F^{\mu e}_{2 \chi^0}$ is given by
\beqn
F^{\mu e}_{2 \chi^0}= \sum_{i=1}^4 \sum_{j=1}^8 \bigg[\frac{-m_{\mu}(m_{\mu} +m_{e})}{192 \pi^2 m^2_{\tilde{\chi_i}^0}}
\{C'^L_{4ij} C'^{L*}_{3ij} + C'^R_{4ij} C'^{R*}_{3ij} \} F_3 \left(\frac{M^2_{\tilde{\tau_j}}}{m^2_{\tilde{\chi_i}^0}}\right)\nonumber\\
- \frac{(m_{\mu} +m_{e})}{64 \pi^2 m_{\tilde{\chi_i}^0}}
\{C'^L_{4ij} C'^{R*}_{3ij} + C'^R_{4ij} C'^{L*}_{3ij} \} F_4 \left(\frac{M^2_{\tilde{\tau_j}}}{m^2_{\tilde{\chi_i}^0}}\right)\bigg]\,,
\label{33}
\eeqn
where
\beq
F_3(x)= \frac{1}{(x-1)^4} \{- x^3 +6x^2 -3x -2 -6x \ln x \}
\label{34}
\eeq
and 
\beq
F_4(x)=\frac{1}{(x-1)^3} \{-x^2 +1 +2x\ln x \}\,.
\label{35}
\eeq
 The contributions  from the W exchange $F^{\mu e}_{2 W}$ is given by  

\beqn
F^{\mu e}_{2 W}= \sum_{i=1}^4 \frac{{m_{\mu}(m_{\mu} +m_{e})}}{32 \pi^2 m^2_W}  [C^W_{Li4} C^{W*}_{Li3} 
+C^W_{Ri4} C^{W*}_{Ri3} ] F_W \left(\frac{m^2_{\psi_i}}{m^2_W}\right)\nonumber\\
 + \frac{{m_{\psi_i}(m_{\mu} +m_{e})}}{32 \pi^2 m^2_W} 
  { [C^W_{Li4} C^{W*}_{Ri3} 
+C^W_{Ri4} C^{W*}_{Li3} ]} G_W \left(\frac{m^2_{\psi_i}}{m^2_W}\right),
\eeqn
where the form factors are given by
\begin{align}
F_{W}(x)&=\frac{1}{6(x-1)^{4}}\left[4 x^4- 49x^{3}+18 x^3 \ln x+78x^{2}-43 x +10 \right]
\end{align}
and
\begin{align}
G_{W}(x)&=\frac{1}{(x-1)^{3}}\left[4 -15 x+12 x^2 - x^3-6 x^2 \ln x \right]\,.
\end{align}
The contribution $F^{\mu e}_{2 Z}$ from the Z exchange is given by

\beqn
F^{\mu e}_{2 Z}=  \sum_{\beta=1}^{4} \frac{{m_{\mu}(m_{\mu} +m_{e})}}{64 \pi^2 m^2_Z}  [C^Z_{L\beta 4} C^{Z*}_{L\beta 3} 
+C^Z_{R\beta 4} C^{Z*}_{R\beta 3} ]  F_Z \left(\frac{m^2_{\tau_{\beta}}}{m^2_Z}\right)\nonumber\\
 + \frac{{m_{\tau_{\beta}}(m_{\mu} +m_{e})}}{64 \pi^2 m^2_Z} 
  { [C^Z_{L\beta 4} C^{Z*}_{R\beta 3} 
+C^Z_{R\beta 4} C^{Z*}_{L\beta 3} ]} G_Z \left(\frac{m^2_{\tau_{\beta}}}{m^2_Z}\right),
\eeqn
where
\begin{align}
F_{Z}(x)&=\frac{1}{3(x-1)^{4}}\left[-5 x^4+14x^{3}-39 x^2+18 x^2 \ln x+38 x -8 \right]
\end{align}
and
\begin{align}
G_{Z}(x)&=\frac{2}{(x-1)^{3}}\left[x^3 + 3 x-6 x \ln x-4 \right].
\end{align}

\noindent
The chargino contribution  $F^{\mu e}_{3 \chi^+}$ is given by
\beqn
F^{\mu e}_{3 \chi^+}= \sum_{i=1}^2 \sum_{j=1}^8 \frac{(m_{\mu} +m_{e})m_{\tilde{\chi_i}^+} }{32 \pi^2 M^2_{\tilde{\nu_j}}}
\{ C^L_{4ij} C^{R*}_{3ij} - C^R_{4ij} C^{L*}_{3ij} \} 
F_5\left(\frac{m^2_{\tilde{\chi_i}^+}}{M^2_{\tilde{\nu_j}}}\right)\,,
\label{36}
\eeqn
where
\beq
F_5(x)= \frac{1}{2(x-1)^2} \{-x +3 + \frac{2\ln x}{1-x} \}\,.
\label{37}
\eeq

\noindent
The neutralino contribution  $F^{\mu e}_{3 \chi^0}$ is given by
\beqn
F^{\mu e}_{3 \chi^0}= \sum_{i=1}^4 \sum_{j=1}^8 \frac{(m_{\mu} +m_{e})m_{\tilde{\chi_i}^0} }{32 \pi^2 M^2_{\tilde{\tau_j}}}
\{ C'^L_{4ij} C'^{R*}_{3ij} - C'^R_{4ij} C'^{L*}_{3ij} \} 
F_6\left(\frac{m^2_{\tilde{\chi_i}^0}}{M^2_{\tilde{\tau_j}}}\right)\,,
\label{38}
\eeqn
where
\beq
F_6(x)=  \frac{1}{2(x-1)^2} \{x +1 + \frac{2 x \ln x}{1-x} \}\,.
\label{39}
\eeq

\noindent
The W boson contrition $F^{\mu e}_{3 W}$ is given by

\begin{align}
F^{\mu e}_{3 W}=- \sum_{i=1}^4 \frac{{m_{\psi_i}(m_{\mu} +m_{e})}}{32 \pi^2 m^2_W}   [C^W_{Li4} C^{W*}_{Ri3} 
-C^W_{Ri4} C^{W*}_{Li3} ] 
I_1\left(\frac{m^{2}_{{\psi}_{i}}}{m^{2}_{W}}\right) \,,
\end{align}
where the functions $C_L^W$ and $C_R^W$ are given in section 4 and the 
 form factor $I_1$  is given by

\begin{align}
I_1(x)&=\frac{2}{(1-x)^{2}}\left[1-\frac{11}{4}x +\frac{1}{4}x^2-\frac{3 x^2\ln x}{2(1-x)} \right]\,.
\end{align}

\noindent
And finally,  the Z exchange diagram contribution $F^{\mu e}_{3 Z}$  is given by

\begin{align}
F^{\mu e}_{3 Z}=   \sum_{\beta=1}^{4}\frac{{(m_{\mu} +m_{e})}}{32 \pi^2}  \frac{m_{\tau_\beta}}{m^2_Z}
[C^Z_{L4 \beta } C^{Z*}_{R3 \beta } 
-C^Z_{R4 \beta } C^{Z*}_{L3 \beta } ] 
I_2\left(\frac{m^{2}_{\tau_{\beta}}}{m^{2}_{Z}}\right) \,,
\end{align}
where the form factor $I_2$  is given by
\begin{align}
I_2(x)&=\frac{2}{(1-x)^{2}}\left[1+\frac{1}{4}x +\frac{1}{4}x^2+\frac{3 x\ln x}{2(1-x)} \right]\,.
\label{23}
\end{align}

\section{Estimate of size of  \b \label{sec6} }

In this section we give a numerical analysis for the branching ratio \b. 
The analysis is done in an MSSM extension with soft breaking parameters taken at the electroweak scale. Thus no renormalization group running of GUT scale parameters is needed. The parameters entering the analysis are summarized in the appendix. The scalar mass and trilinear coupling parameters are $m_0$ and $A_0$ in the slepton mass squared matrix. The corresponding ones in the sneutrino mass squared matrix are $m_0^{\tilde{\nu }}$ and $A_0^{\tilde{\nu }}$ where
\begin{align}
m_0^2 &=\tilde M^2_{\tau L}=\tilde M^2_E =\tilde M^2_{\tau}=\tilde M^2_{\chi}=\tilde M^2_{\mu L} =\tilde M^2_{\mu}=\tilde M^2_{e L} =\tilde M^2_{e}\nonumber\\
m_0^{\tilde{\nu }} &=\tilde M^2_N=\tilde M^2_{\nu_\tau} =\tilde M^2_{\nu_\mu} =\tilde M^2_{\nu_e}\nonumber\\
A_0&=A_{\tau}=A_{\mu}=A_e=A_E\nonumber\\
A_0^{\tilde{\nu }}&=A_{\nu_\tau}=A_{\nu_\mu}=A_{\nu_e}=A_N
\end{align}
The branching ratio \b arises as a consequence of mixing induced by the parameters $f_3, f_3', f_3''$ and 
$f_4, f_4', f_4''$ where $f$'s are complex parameters and their arguments are the CP violating phases.
The branching ratio \b is a sensitive function of both the magnitudes as well as the phases of the mixing parameters
$f$. We discuss the dependence of \b on these below.

 Fig. \ref{fig2} exhibits the variation of \b as a function of the CP violating phases $\chi_3'$ (left panel)
 and $\chi_3''$ (right panel).
As the two panels of Fig.\ref{fig2} show  \b is a sensitive function of these phases and can vary by an order of magnitude or more as the phases vary.  The solid horizontal line gives the current experimental upper limit on \b from the MEG 
experiment~\cite{Adam:2011ch}.
A very similar analysis holds when we vary 
the CP phases $\chi_4$ and $\chi_4'$ as exhibited in Fig.\ref{fig3}.
Figure \ref{fig4} gives the relative strength of the magnetic and the electric dipole transition operators to \b. Thus 
the left panel of Fig.\ref{fig4} exhibits the relative strength of the contributions from the magnetic dipole operator and the electric dipole moment operator as a function of the CP phase $\chi_4''$. The right panel of Fig.\ref{fig4} exhibits the dependence of
the electron EDM $d_e$ as a function of $\chi_4''$ where the solid horizontal line gives the current upper limit
on $d_e$ from the ACME Collaboration~\cite{Baron:2013eja}. 
 Thus the right panel delineates the allowed regions of the parameter space, i.e., regions  consistent with 
the experimental upper limit constraint on $d_e$. 
The left panel of Fig.\ref{fig5} exhibits the variation of  \b as a function of $\chi''_3$ for different values of the mixing parameter $|f''_3|$ 
while  the right panel of Fig.\ref{fig5} gives the electric dipole moment of the electron 
with the horizontal solid line giving the  experimental upper limit on it. 
 A comparison of the left and the right panels show
the regions of $\chi_3''$ consistent with the current experimental upper limits on \b and on $d_e$ and accessible with reasonable
improvement in the sensitivity of  experiment in the future. 

In table1  we illustrate numerically the relative strengths of the magnetic and the electric  dipole transition operators
to \b. Here we also show that the analysis is consistent with the current upper limits on the ${\cal B}(\tau\to \mu\gamma)$,
on $d_e$ and the data on the neutrino masses. Thus  the current experimental limit on ${\cal B}(\tau\to \mu\gamma)$ is
${\cal B}(\tau\to \mu\gamma) < 4.4 \times 10^{-8}$ (BaBar)~\cite{Aubert:2009ag}
 and ${\cal B}(\tau\to \mu\gamma) < 4.5 \times 10^{-8}$ (Belle)~\cite{Hayasaka:2007vc}. 
Since the theoretical prediction of ${\cal B}(\tau\to \mu\gamma)$ in this case is smaller by several orders of magnitude than
the current experimental limit this decay mode is not of imminent interest in this case.
In  table 2 we give a numerical analysis of the form factors $F_2$ and $F_3$ and their sub pieces arising from 
the supersymmetric
and the non-supersymmetric loops. Also listed are ${\cal B}_m$ and ${\cal B}_e$ as well as $d_e$ and the neutrino masses. 
One finds that typically the magnetic dipole contributions dominate the electric dipole contributions.
The neutrino mass results of  table 1 and  table 2 are consistent 
with the constraint on the sum of the neutrino masses from cosmology, i.e., $\sum_{i} m_{\nu_i} <0.44$eV (95\% CL)~\cite{Hinshaw:2012aka} and with the data 
on 
neutrino oscillations which give the neutrino mass squared differences so that~\cite{Schwetz:2008er}
\begin{gather}
\label{6.1b}
\Delta m^2_{31}\equiv m_3^2-m_1^2= 2.4^{+0.12}_{-0.11} \times 10^{-3} ~eV^2  \ , \\
\Delta m_{21}^2\equiv m_2^2- m_1^2= 7.65^{+0.23}_{-0.20} \times 10^{-5}~eV^2. 
\label{6.1c}
\end{gather}
Fig. (\ref{fig6}) exhibits a variation of \b as a function of the mirror masses $m_E$ and $m_N$. All points of the four curves of Fig. (6) are consistent with the constraints set on the neutrino masses by \cref{6.1b} and (\ref{6.1c}).
We note that in the analysis of Fig. (2)- Fig.(4) the mass parameters are typically low. For instance in the analysis of Fig.4 we have used $|\mu|=310$ GeV, $M_1=180$ GeV and $M_2=140$ GeV 
and our vectorlike masses are chosen so that
$m_E=m_N=150, 200, 250, 300$.  Such a choice may be close to the LHC exclusion plots based on LHC
RUN I data and could be close to 
being probed with more data. We should note that
the LHC particle searches are very model dependent as can be seen from the analyses of 
\cite{Aad:2014vma,Aad:2014nua}. For instance in the ATLAS analysis of \cite{Aad:2014nua}
the lightest chargino mass is excluded up to 700 GeV, 380 GeV, 345 GeV or 148 GeV for a massless 
neutralino depending on the allowed decay channels.  These results would be even more model
dependent if the neutralino is assumed massive with a varying mass.  Thus while the current limits 
from LHC do not directly apply to our analysis, the choice of low mass parameters point to the possibility 
that they could be probed in RUN II of the LHC. It would thus be very interesting to carryout a 
signal analysis of this model  specifically, for instance, for multilepton searches.  Such an analysis, however,
is beyond the scope of this paper.

\section{Conclusion\label{sec7}}
In this work we have given an analysis of $\mu \to e \gamma$ decay with inclusion of a vectorlike 
leptonic generation where mixings appear between  leptons and mirror leptons as well as between
sleptons and mirror sleptons. The decay $\mu \to e \gamma$ arises from diagrams with charginos
and sneutrinos and mirror sneutrinos,  and neutralinos, sleptons and mirror sleptons in the loops.
Additionally electroweak loops are included where W, Z and leptons and mirror leptons and neutrinos are exchanged.
An analytic analysis of these contributions is given in section 5 while a detailed numerical analysis is 
given in section 6. Here it is shown that the current experimental limits from the MEG experiment
put constraints on the parameter space of models. Further, the size of the new contributions are 
such that improvement in experiment will either reveal new physics or the improved experimental
results will be able to probe large parts of the parameter space of the extended MSSM model. 
Thus the MEG experiment is continuing to collect data and is expected
to explore the $\mu\to e+\gamma$  decay down to a branching
ratio sensitivity of a few times $10^{-13}$  in the next few
years. This will allow a further probe of this new class of MSSM extensions.\\

\noindent
{\bf Acknowledgments:}
PN's research  is  supported in part by the NSF grant PHY-1314774.\\

{\section{Appendix: The scalar mass squared matrices\label{sec8}}
   For convenience we collect here all the contributions to the scalar mass squared matrices
   arising from the F terms.  They are given by
\beq
{\cal L}^{\rm mass}_F= {\cal L}_C^{\rm mass} +{\cal L}_N^{\rm mass}\ ,
\eeq
where  ${\cal L}_C^{\rm mass}$ gives the mass terms for the charged sleptons while
$ {\cal L}_N^{mass}$ gives the mass terms for the  sneutrinos. For ${\cal L}_C^{\rm mass}$ we have
\begin{gather}
-{\cal L}_C^{\rm mass} =\left(\frac{v^2_2 |f'_2|^2}{2} +|f_3|^2+|f_3'|^2+|f_3''|^2\right)\tilde E_R \tilde E^*_R
%\nonumber\\
+\left(\frac{v^2_2 |f'_2|^2}{2} +|f_4|^2+|f_4'|^2+|f_4''|^2\right)\tilde E_L \tilde E^*_L\nonumber\\
+\left(\frac{v^2_1 |f_1|^2}{2} +|f_4|^2\right)\tilde \tau_R \tilde \tau^*_R
+\left(\frac{v^2_1 |f_1|^2}{2} +|f_3|^2\right)\tilde \tau_L \tilde \tau^*_L
+\left(\frac{v^2_1 |h_1|^2}{2} +|f_4'|^2\right)\tilde \mu_R \tilde \mu^*_R\nonumber\\
+\left(\frac{v^2_1 |h_1|^2}{2} +|f_3'|^2\right)\tilde \mu_L \tilde \mu^*_L
+\left(\frac{v^2_1 |h_2|^2}{2} +|f_4''|^2\right)\tilde e_R \tilde e^*_R
+\left(\frac{v^2_1 |h_2|^2}{2} +|f_3''|^2\right)\tilde e_L \tilde e^*_L\nonumber\\
+\Bigg\{-\frac{f_1 \mu^* v_2}{\sqrt{2}} \tilde \tau_L \tilde \tau^*_R
-\frac{h_1 \mu^* v_2}{\sqrt{2}} \tilde \mu_L \tilde \mu^*_R
 -\frac{f'_2 \mu^* v_1}{\sqrt{2}} \tilde E_L \tilde E^*_R
+\left(\frac{f'_2 v_2 f^*_3}{\sqrt{2}}  +\frac{f_4 v_1 f^*_1}{\sqrt{2}}\right) \tilde E_L \tilde \tau^*_L\nonumber\\
+\left(\frac{f_4 v_2 f'^*_2}{\sqrt{2}}  +\frac{f_1 v_1 f^*_3}{\sqrt{2}}\right) \tilde E_R \tilde \tau^*_R
+\left(\frac{f'_3 v_2 f'^*_2}{\sqrt{2}}  +\frac{h_1 v_1 f'^*_4}{\sqrt{2}}\right) \tilde E_L \tilde \mu^*_L
+\left(\frac{f'_2 v_2 f'^*_4}{\sqrt{2}}  +\frac{f'_3 v_1 h^*_1}{\sqrt{2}}\right) \tilde E_R \tilde \mu^*_R\nonumber\\
+\left(\frac{f''^*_3 v_2 f'_2}{\sqrt{2}}  +\frac{f''_4 v_1 h^*_2}{\sqrt{2}}\right) \tilde E_L \tilde e^*_L
+\left(\frac{f''_4 v_2 f'^*_2}{\sqrt{2}}  +\frac{f''^*_3 v_1 h^*_2}{\sqrt{2}}\right) \tilde E_R \tilde e^*_R
+f'_3 f^*_3 \tilde \mu_L \tilde \tau^*_L +f_4 f'^*_4 \tilde \mu_R \tilde \tau^*_R\nonumber\\
+f_4 f''^*_4 \tilde {e}_R \tilde{\tau}^*_R
+f''_3 f^*_3 \tilde {e}_L \tilde{\tau}^*_L
+f''_3 f'^*_3 \tilde {e}_L \tilde{\mu}^*_L
+f'_4 f''^*_4 \tilde {e}_R \tilde{\mu}^*_R
-\frac{h_2 \mu^* v_2}{\sqrt{2}} \tilde{e}_L \tilde{e}^*_R
+H.c. \Bigg\}\,.
%\label{11a}
\end{gather}
We define the scalar mass squared   matrix $M^2_{\tilde \tau}$  in the basis $(\tilde  \tau_L, \tilde E_L, \tilde \tau_R,
\tilde E_R, \tilde \mu_L, \tilde \mu_R, \tilde e_L, \tilde e_R)$. We  label the matrix  elements of these as $(M^2_{\tilde \tau})_{ij}= M^2_{ij}$ where the elements of the matrix are given by
\begin{align}
M^2_{11}&=\tilde M^2_{\tau L} +\frac{v^2_1|f_1|^2}{2} +|f_3|^2 -m^2_Z \cos 2 \beta \left(\frac{1}{2}-\sin^2\theta_W\right), \nonumber\\
M^2_{22}&=\tilde M^2_E +\frac{v^2_2|f'_2|^2}{2}+|f_4|^2 +|f'_4|^2+|f''_4|^2 +m^2_Z \cos 2 \beta \sin^2\theta_W, \nonumber\\
M^2_{33}&=\tilde M^2_{\tau} +\frac{v^2_1|f_1|^2}{2} +|f_4|^2 -m^2_Z \cos 2 \beta \sin^2\theta_W, \nonumber\\
M^2_{44}&=\tilde M^2_{\chi} +\frac{v^2_2|f'_2|^2}{2} +|f_3|^2 +|f'_3|^2+|f''_3|^2 +m^2_Z \cos 2 \beta \left(\frac{1}{2}-\sin^2\theta_W\right), \nonumber
\end{align}
\begin{align}
M^2_{55}&=\tilde M^2_{\mu L} +\frac{v^2_1|h_1|^2}{2} +|f'_3|^2 -m^2_Z \cos 2 \beta \left(\frac{1}{2}-\sin^2\theta_W\right), \nonumber\\
M^2_{66}&=\tilde M^2_{\mu} +\frac{v^2_1|h_1|^2}{2}+|f'_4|^2 -m^2_Z \cos 2 \beta \sin^2\theta_W, \nonumber\\
M^2_{77}&=\tilde M^2_{e L} +\frac{v^2_1|h_2|^2}{2}+|f''_3|^2 -m^2_Z \cos 2 \beta \left(\frac{1}{2}-\sin^2\theta_W\right), \nonumber\\
M^2_{88}&=\tilde M^2_{e} +\frac{v^2_1|h_2|^2}{2}+|f''_4|^2 -m^2_Z \cos 2 \beta \sin^2\theta_W\ . \nonumber
\end{align}

\begin{align}
M^2_{12}&=M^{2*}_{21}=\frac{ v_2 f'_2f^*_3}{\sqrt{2}} +\frac{ v_1 f_4 f^*_1}{\sqrt{2}} ,
M^2_{13}=M^{2*}_{31}=\frac{f^*_1}{\sqrt{2}}(v_1 A^*_{\tau} -\mu v_2),
M^2_{14}=M^{2*}_{41}=0,\nonumber\\
 M^2_{15} &=M^{2*}_{51}=f'_3 f^*_3,
 M^{2*}_{16}= M^{2*}_{61}=0,  M^{2*}_{17}= M^{2*}_{71}=f''_3 f^*_3,  M^{2*}_{18}= M^{2*}_{81}=0,\nonumber\\
M^2_{23}&=M^{2*}_{32}=0,
M^2_{24}=M^{2*}_{42}=\frac{f'^*_2}{\sqrt{2}}(v_2 A^*_{E} -\mu v_1),  M^2_{25} = M^{2*}_{52}= \frac{ v_2 f'_3f'^*_2}{\sqrt{2}} +\frac{ v_1 h_1 f^*_4}{\sqrt{2}} ,\nonumber\\
 M^2_{26} &=M^{2*}_{62}=0,  M^2_{27} =M^{2*}_{72}=  \frac{ v_2 f''_3f'^*_2}{\sqrt{2}} +\frac{ v_1 h_2 f''^*_4}{\sqrt{2}},  M^2_{28} =M^{2*}_{82}=0, \nonumber\\
M^2_{34}&=M^{2*}_{43}= \frac{ v_2 f_4 f'^*_2}{\sqrt{2}} +\frac{ v_1 f_1 f^*_3}{\sqrt{2}}, M^2_{35} =M^{2*}_{53} =0, M^2_{36} =M^{2*}_{63}=f_4 f'^*_4,\nonumber\\
 M^2_{37} &=M^{2*}_{73} =0,  M^2_{38} =M^{2*}_{83} =f_4 f''^*_4,\nonumber\\
M^2_{45}&=M^{2*}_{54}=0, M^2_{46}=M^{2*}_{64}=\frac{ v_2 f'_2 f'^*_4}{\sqrt{2}} +\frac{ v_1 f'_3 h^*_1}{\sqrt{2}}, \nonumber\\
 M^2_{47} &=M^{2*}_{74}=0,  M^2_{48} =M^{2*}_{84}=  \frac{ v_2 f'_2f''^*_4}{\sqrt{2}} +\frac{ v_1 f''_3 h^*_2}{\sqrt{2}},\nonumber\\
M^2_{56}&=M^{2*}_{65}=\frac{h^*_1}{\sqrt{2}}(v_1 A^*_{\mu} -\mu v_2),
 M^2_{57} =M^{2*}_{75}=f''_3 f'^*_3,  \nonumber\\
 M^2_{58} &=M^{2*}_{85}=0,  M^2_{67} =M^{2*}_{76}=0,\nonumber\\
 M^2_{68} &=M^{2*}_{86}=f'_4 f''^*_4,  M^2_{78}=M^{2*}_{87}=\frac{h^*_2}{\sqrt{2}}(v_1 A^*_{e} -\mu v_2)\ . \nonumber
\label{14}
\end{align}

We can diagonalize this hermitian mass squared  matrix  by the
 unitary transformation
\begin{gather}
 \tilde D^{\tau \dagger} M^2_{\tilde \tau} \tilde D^{\tau} = diag (M^2_{\tilde \tau_1},
M^2_{\tilde \tau_2}, M^2_{\tilde \tau_3},  M^2_{\tilde \tau_4},  M^2_{\tilde \tau_5},  M^2_{\tilde \tau_6},  M^2_{\tilde \tau_7},  M^2_{\tilde \tau_8} )\ .
\end{gather}

%%%%%%%%%%%%%%%%%%%%%%%%%%%%%%%%%%%%%%%%%%%%%%%%%%%%%%%%

For ${\cal L}_N^{\rm mass}$ we have
\begin{multline}
-{\cal L}_N^{\rm mass}=
\left(\frac{v^2_1 |f_2|^2}{2}
 +|f_3|^2+|f_3'|^2+|f_3''|^2\right)\tilde N_R \tilde N^*_R\\
 +\left(\frac{v^2_1 |f_2|^2}{2}+|f_5|^2+|f_5'|^2+|f_5''|^2\right)\tilde N_L \tilde N^*_L
+\left(\frac{v^2_2 |f'_1|^2}{2}+|f_5|^2\right)\tilde \nu_{\tau R} \tilde \nu^*_{\tau R}\\
+\left(\frac{v^2_2 |f'_1|^2}{2}
+|f_3|^2\right)\tilde \nu_{\tau L} \tilde \nu^*_{\tau L}
+\left(\frac{v^2_2 |h'_1|^2}{2}
+|f_3'|^2\right)\tilde \nu_{\mu L} \tilde \nu^*_{\mu L}
+\left(\frac{v^2_2 |h'_1|^2}{2}
+|f_5'|^2\right)\tilde \nu_{\mu R} \tilde \nu^*_{\mu R}\nonumber\\
+\left(\frac{v^2_2 |h'_2|^2}{2}
+|f_3''|^2\right)\tilde \nu_{e L} \tilde \nu^*_{e L}
+\left(\frac{v^2_2 |h'_2|^2}{2}
+|f_5''|^2\right)\tilde \nu_{e R} \tilde \nu^*_{e R}\nonumber\\
+\Bigg\{ -\frac{f_2 \mu^* v_2}{\sqrt{2}} \tilde N_L \tilde N^*_R
-\frac{f'_1 \mu^* v_1}{\sqrt{2}} \tilde \nu_{\tau L} \tilde \nu^*_{\tau R}
-\frac{h'_1 \mu^* v_1}{\sqrt{2}} \tilde \nu_{\mu L} \tilde \nu^*_{\mu R}
+\left(\frac{f_5 v_2 f'^*_1}{\sqrt{2}}  -\frac{f_2 v_1 f^*_3}{\sqrt{2}}\right) \tilde N_L \tilde \nu^*_{\tau L}\nonumber\\
+\left(\frac{f_5 v_1 f^*_2}{\sqrt{2}}  -\frac{f'_1 v_2 f^*_3}{\sqrt{2}}\right) \tilde N_R \tilde \nu^*_{\tau R}
+\left(\frac{h'_1 v_2 f'^*_5}{\sqrt{2}}  -\frac{f'_3 v_1 f^*_2}{\sqrt{2}}\right) \tilde N_L \tilde \nu^*_{\mu L}
+\left(\frac{f''_5 v_1 f^*_2}{\sqrt{2}}  -\frac{f''^*_3 v_2 h'_2}{\sqrt{2}}\right) \tilde N_R \tilde \nu^*_{e R}\nonumber\\
+\left(\frac{h'^*_2 v_2 f''_5}{\sqrt{2}}  -\frac{f''^*_3 v_1 f_2}{\sqrt{2}}\right) \tilde N_L \tilde \nu^*_{e L}
+\left(\frac{f'_5 v_1 f^*_2}{\sqrt{2}}  -\frac{h'_1 v_2 f'^*_3}{\sqrt{2}}\right) \tilde N_R \tilde \nu^*_{\mu R}\nonumber\\
+f'_3 f^*_3 \tilde \nu_{\mu L} \tilde \nu_{\tau^*_L} +f_5 f'^*_5 \tilde \nu_{\mu R} \tilde \nu^*_{\tau R}
-\frac{h'_2 \mu^* v_1}{\sqrt{2}} \tilde{\nu}_{e L} \tilde{\nu}^*_{e R}\\
+f''_3 f^*_3   \tilde{\nu}_{e L} \tilde{\nu}^*_{\tau L}
+f_5 f''^*_5   \tilde{\nu}_{e R} \tilde{\nu}^*_{\tau R}
+f''_3 f'^*_3   \tilde{\nu}_{e L} \tilde{\nu}^*_{\mu L}
+f'_5 f''^*_5   \tilde{\nu}_{e R} \tilde{\nu}^*_{\mu R}
+H.c. \Bigg\}.
\label{11b}
\end{multline}

 Next we write the   mass squared  matrix in the sneutrino sector the basis $(\tilde  \nu_{\tau L}, \tilde N_L,$
$ \tilde \nu_{\tau R}, \tilde N_R, \tilde  \nu_{\mu L},\tilde \nu_{\mu R}, \tilde \nu_{e L}, \tilde \nu_{e R} )$.
 Thus here we denote the sneutrino mass squared matrix in the form
$(M^2_{\tilde\nu})_{ij}=m^2_{ij}$ where

\begin{align}
m^2_{11}&=\tilde M^2_{\tau L} +\frac{v^2_2|f'_1|^2}{2} +|f_3|^2 +\frac{1}{2}m^2_Z \cos 2 \beta,  \nonumber\\
m^2_{22}&=\tilde M^2_N +\frac{v^2_1|f_2|^2}{2} +|f_5|^2 +|f'_5|^2+|f''_5|^2, \nonumber\\
m^2_{33}&=\tilde M^2_{\nu_\tau} +\frac{v^2_2|h'_1|^2}{2} +|f_5|^2,  \nonumber\\
m^2_{44}&=\tilde M^2_{\chi} +\frac{v^2_1|f_2|^2}{2} +|f_3|^2 +|f'_3|^2+|f''_3|^2 -\frac{1}{2}m^2_Z \cos 2 \beta, \nonumber\\
m^2_{55}&=\tilde M^2_{\mu L} +\frac{v^2_2|f'_1|^2}{2}+|f'_3|^2 +\frac{1}{2}m^2_Z \cos 2 \beta,  \nonumber\\
m^2_{66}&=\tilde M^2_{\nu_\mu} +\frac{v^2_2|h'_1|^2}{2} +|f'_5|^2,  \nonumber\\
m^2_{77}&=\tilde M^2_{e L} +\frac{v^2_2|h'_2|^2}{2} +|f''_3|^2+\frac{1}{2}m^2_Z \cos 2 \beta,  \nonumber\\
m^2_{88}&=\tilde M^2_{\nu_e} +\frac{v^2_2|h'_2|^2}{2}  +|f''_5|^2,  \nonumber
\end{align}

\begin{align}
m^2_{12}&=m^{2*}_{21}=\frac{v_2 f_5 f'^*_1}{\sqrt{2}}-\frac{ v_1 f_2 f^*_3}{\sqrt{2}},
~m^2_{13}=m^{2*}_{31}=\frac{f'^*_1}{\sqrt{2}}(v_2 A^*_{\nu_\tau} -\mu v_1)\nonumber\,,\\
m^2_{14}&=m^{2*}_{41}=0,
~m^2_{15}=m^{2*}_{51}= f'_3 f^*_3, m^2_{16}=m^{2*}_{61}=0,\nonumber\\
m^2_{17}&=m^{2*}_{71}= f''_3 f^*_3, m^2_{18}=m^{2*}_{81}=0,\nonumber\\
m^2_{23}&=m^{2*}_{32}=0,
m^2_{24}=m^{2*}_{42}=\frac{f^*_2}{\sqrt{2}}(v_{1}A^*_N-\mu v_2), \nonumber\\
m^2_{25}&=m^{2*}_{52}=-\frac{v_{1}f^*_2 f'_3}{\sqrt{2}}+\frac{h'_1 v_2 f'^*_5}{\sqrt{2}},\nonumber\\
m^2_{26}&=m^{2*}_{62}=0, m^2_{27}=m^{2*}_{72}=-\frac{v_{1}f^*_2 f''_3}{\sqrt{2}}+\frac{h'_2 v_2 f''^*_5}{\sqrt{2}}\,,
\nonumber\\ 
m^2_{28}&=m^{2*}_{82}=0, m^2_{34}=m^{2*}_{43}=\frac{v_1 f^*_2 f_5}{\sqrt{2}}-\frac{v_2 f'_1 f^*_3}{\sqrt{2}},\nonumber\\
%\end{align}
m^2_{35}&=m^{2*}_{53}=0, m^2_{36}=m^{2*}_{63}=f_5 f'^*_5, \nonumber\\
m^2_{37}&=m^{2*}_{73}=0, m^2_{38}=m^{2*}_{83}=f_5 f''^*_5, \nonumber\\
m^2_{45}&=m^{2*}_{54}=0, m^2_{46}=m^{2*}_{64}=-\frac{h'^*_1 v_2 f'_3}{\sqrt{2}}+\frac{v_1 f_2 f'^*_5}{\sqrt{2}}, 
\nonumber\\
%\end{align}
m^2_{47}&=m^{2*}_{74}=0, 
m^2_{48}=m^{2*}_{84}=\frac{v_1 f_2 f''^*_5}{\sqrt{2}}-\frac{v_2 h'^*_2 f''_3}{\sqrt{2}},\nonumber\\
 m^2_{56}&=m^{2*}_{65}=\frac{h'^*_1}{\sqrt{2}}(v_2 A^*_{\nu_\mu}-\mu v_1), \nonumber\\
m^2_{57}&=m^{2*}_{75}= f''_3 f'^*_3, m^2_{58}=m^{2*}_{85}=0, \nonumber\\
m^2_{67}&=m^{2*}_{76}=0, m^2_{68}=m^{2*}_{86}= f'_5 f''^*_5, \nonumber\\
m^2_{78}&=m^{2*}_{87}=\frac{h'^*_2}{\sqrt{2}}(v_2 A^*_{\nu_e}-\mu v_1).  
%\label{15}
\end{align}

We can diagonalize the sneutrino mass square matrix  by the  unitary transformation 
\begin{equation}
 \tilde D^{\nu\dagger} M^2_{\tilde \nu} \tilde D^{\nu} = \text{diag} (M^2_{\tilde \nu_1}, M^2_{\tilde \nu_2}, M^2_{\tilde \nu_3},  M^2_{\tilde \nu_4},M^2_{\tilde \nu_5},  M^2_{\tilde \nu_6}, M^2_{\tilde \nu_7}, M^2_{\tilde \nu_8})\ .
\end{equation}

 \begin{figure}
      \includegraphics[scale=.39]{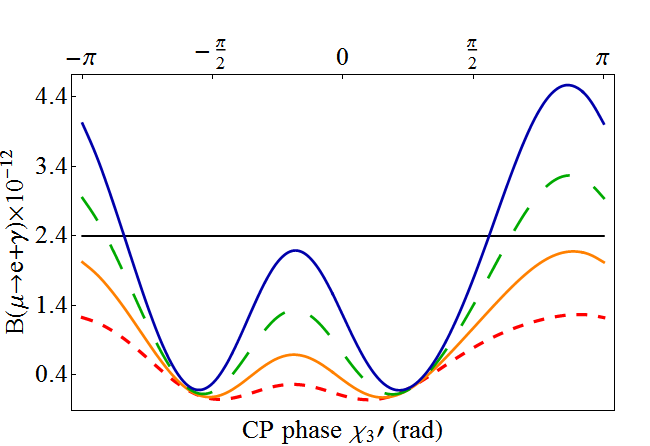}
 \includegraphics[scale=.39]{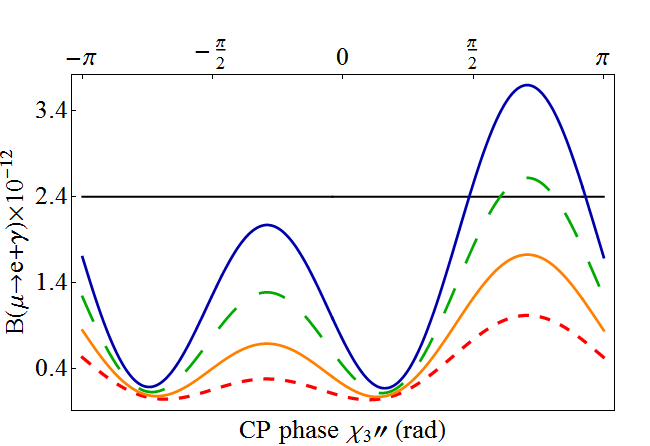}
          \caption{
Left Panel:  An exhibition of the dependence of \b  on $\chi_3^{\prime}$ where $\chi_3= \chi_3^{\prime\prime}= \chi_3^{\prime}$.
The \b curves (bottom to top at $\chi_3'=\pi$)
 are for the cases 
 when $|f_3| = |f_3^{\prime}| = |f_3^{\prime\prime}| = 5\times 10^{-5}$, $7\times 10^{-5}$,
$9\times 10^{-5}$, and $11\times 10^{-5}$. 
 Right Panel: An exhibition of \b as a function of the  $\chi_3''$  where the curves (bottom to top at $\chi_3''=\pi$) are for  
 the cases
  $|f_3| = 5\times 10^{-5}$, $7\times 10^{-5}$, $9\times 10^{-5}$, $11\times 10^{-5}$
  where $|f_3|=|f_3'|=5\times10^{-5}$  and $\chi_3=\chi_3'=0.3$. The common parameters for both panels  are $\tb=5$, $|\mu|=500$, $|M_1|=130$, $|M_2|=110$, $m_N=260$, $m_E=280$, $m_0=4\times10^{5}$, $m_0^{\tilde{\nu }}=5\times10^5$, $|A_0|= |A_0^{\tilde{\nu }}|= 6\times10^{5}$, $\alpha _1=0.4$, $\alpha _2=\alpha _{\mu }=0.5$, $\alpha _{A_0}= \alpha _{A_0^{\tilde{\nu }}}=1$, $|f_4|=|f_4'|=1$, $|f_4''|=0.1$, $|f_5|=3\times10^{-6}$, $|f_5'|=8\times10^{-7}$, $|f_5''|=5\times10^{-6}$, $\chi_4=1$, $\chi_4'=\chi_4''=0.5$, $\chi_5=\chi_5'=\chi_5''=1$.
  The solid horizontal line is the upper limit from the MEG experiment~\cite{Adam:2011ch}.  
  Here and in the rest of the figures and in the tables all masses are in GeV and phase angles in radian.           }
\label{fig2}
\end{figure}

 \begin{figure}
      \includegraphics[scale=.39]{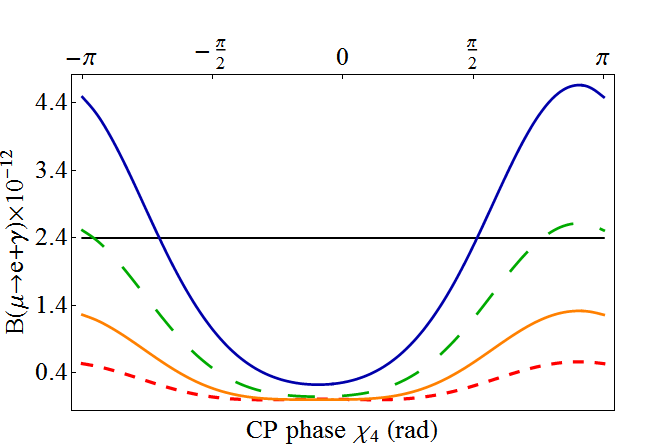}
 \includegraphics[scale=.39]{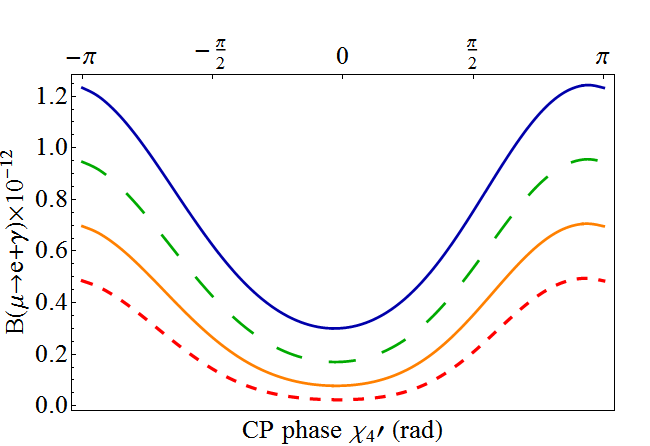}
          \caption{Left Panel: Plot of \b as a function of $\chi_4$ when
          $\chi_4'=\chi_4''=\chi_4$. The curves from bottom to top at $\chi_4=\pi$ are for the cases $|f_4|=|f_4'|=|f_4''|=$ 0.1 , 0.5, 1, and 1.5
           and  $|f_3|=|f_3'|=|f_3''|=5\times10^{-5}$ and $\chi_3= \chi_3^{\prime} = \chi_3^{\prime\prime}=0.3$.     
           The solid horizontal line is the upper limit from the MEG experiment~\cite{Adam:2011ch}.                   
              Right Panel: An exhibition of the dependence of \b on $\chi_4'$.  
              The curves from bottom to top at $\chi_4^{''}=\pi$ are for the cases                       
              when $|f_4^{\prime}|$=0.3, 0.4, 0.5, and 0.6
              and   $|f_4|=|f_4''|=0.4, \chi_4=\chi_4^{\prime\prime}=1$. All other parameters in both panels are the same as in Fig.\ref{fig2}.
 }           

\label{fig3}
\end{figure}

\begin{figure}
   \includegraphics[scale=.39]{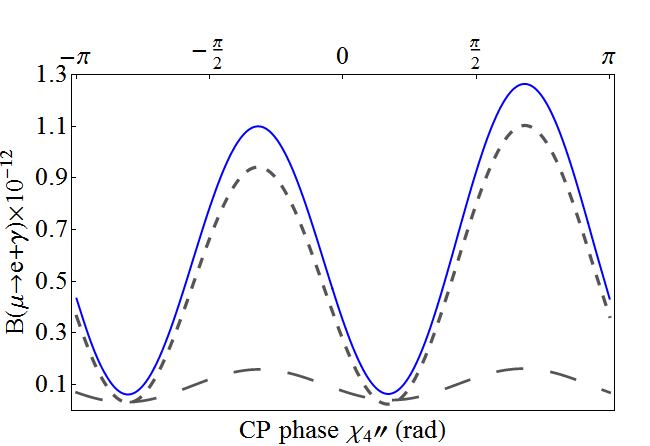}
      \includegraphics[scale=.39]{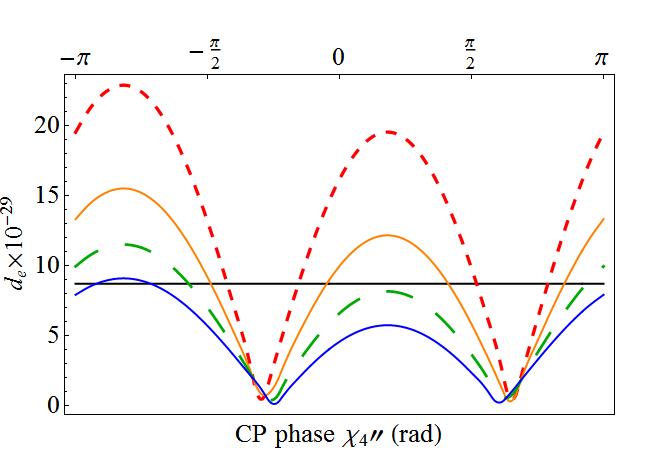}
      \caption{Left Panel:  An exhibition of the magnetic contribution ${\cal B}_m$ (dotted) as given by Eq. (\ref{bm}), the electric contribution 
       ${\cal B}_e$ (dashed) as given by  Eq. (\ref{be}) to \b where the solid curve stands for their sum as a function of $\chi_4^{\prime\prime}$ when
        $m_E=m_N=300$. 
        Right Panel: An exhibition of $d_e$ as a function of $\chi_4^{\prime\prime}$.
        The curves from top to bottom  at $\chi_4^{''}=\pi$ are 
         for values of $m_E = m_N = 150$, 200, 250, and 300. 
         The solid horizontal line is the upper limit on $d_e$ from the ACME Collaboration~\cite{Baron:2013eja}.         
         The common parameters for the two panels are  $|\mu|=310$, $|M_1|=180$, $|M_2|=140$, $\tan\beta=20$, $m_0=4\times10^{5}$, $|A_0|= 1.5\times10^{6}$, $m_0^{\tilde{\nu }}=4\times10^5$, $|A_0^{\tilde{\nu }}|= 5.1\times10^{6}$, $\alpha _1=0.4$, $\alpha _2=0.2$, $\alpha _{\mu }=0.7$, $\alpha _{A_0}= \alpha _{A_0^{\tilde{\nu }}}=1$. 
The mixings are $|f_3|=7\times10^{-4}$, $|f_3^{\prime}|=1\times10^{-4}$, $|f_3^{\prime\prime}|=2\times10^{-4}$, $|f_4|=3\times10^{-2}$, $|f_4'|=0.4$, $|f_4''|=5\times10^{-2}$, $|f_5|=3.8\times10^{-6}$, 
$|f_5'|=2.2\times10^{-6}$, $|f_5''|=3\times10^{-6}$, $\chi_3=\chi_3'=\chi_3''=1$, $\chi_4=0.3$, $\chi_4'=0.2$, $\chi_5=\chi_5'=\chi_5''=0.5$.
}

\label{fig4}
\end{figure}

\begin{table}\centering
\begin{tabular}{lcc}

 & (i) $m_E=m_N=300$  & \hspace{0.4cm} (ii) $m_E=m_N=150$ \\  \hline
${\cal{B}}_m$  & $1.1\times10^{-12}$ & \hspace{0.5cm}  $8.3\times10^{-12}$  \\
${\cal{B}}_e$  & $1.6\times10^{-13}$ & \hspace{0.5cm}  $1.2\times10^{-12}$  \\ \hline\hline
\b & $1.2\times10^{-12}$ & \hspace{0.5cm} $9.4\times10^{-12}$  \\ 
\bb & $4.8\times10^{-25}$ & \hspace{0.5cm} $8.0\times10^{-25}$  \\ 
$d_e$ ($e$cm) & $6.3\times10^{-30}$ & \hspace{0.5cm} $1.3\times10^{-29}$  \\ 
\hline
$m_{\nu3}$ & $5.0\times10^{-11}$ & \hspace{0.5cm} $5.5\times10^{-11}$ \\
$m_{\nu2}$ & $8.9\times10^{-12}$ & \hspace{0.5cm} $9.6\times10^{-12}$ \\
$m_{\nu1}$ & $ 1.1\times10^{-12}$ & \hspace{0.5cm} $2.5\times10^{-12}$ \\ 
%\hline
\hline\hline
\end{tabular}
\caption{
An exhibition of the numerical values of ${\cal B}_m$, ${\cal B}_e$,
 \b and ${\cal B}(\tau\to \mu+\gamma)$ when
$\chi_4^{\prime\prime}=2$ for two values of $m_E=m_N$ while other parameters are the same as in Fig.\ref{fig4}. 
The values of the electron EDM $d_e$ and the neutrino masses $m_{\nu_1}, m_{\nu_2}, m_{\nu_3}$ are also exhibited. }
\label{table1}
\end{table}

 \begin{figure}
    \includegraphics[scale=.39]{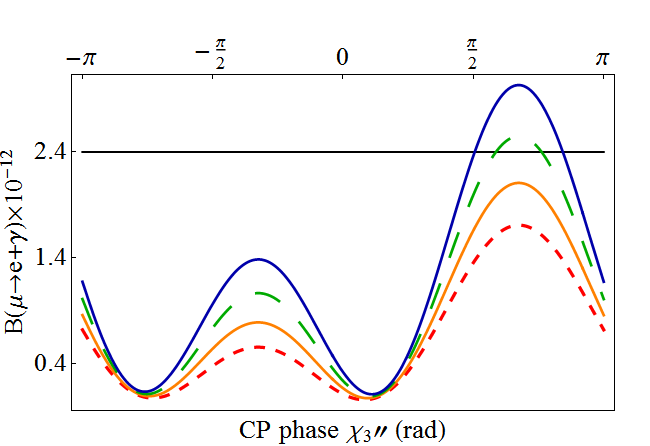}
      \includegraphics[scale=.39]{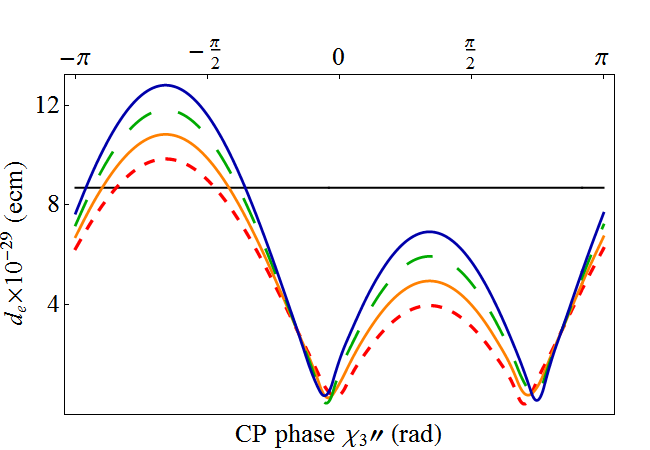}
       \caption{Left Panel: A plot of  \b as a function of  $\chi_3''$ where the curves from bottom to top  at $\chi_3''=2$ are 
       for values of $|f_3''|=7\times10^{-5}$, $8\times10^{-5}$, $9\times10^{-5}$, $1\times10^{-4}$.
         The solid horizontal line is the upper limit from the MEG experiment~\cite{Adam:2011ch}.        
             Right Panel:  Plot of  $d_e$ as a function of  $\chi_3''$ for the same values of $f_3''$ as in the left panel. 
              The solid horizontal line is the upper limit on $d_e$ from the ACME Collaboration~\cite{Baron:2013eja}.              
              The parameters  used are $\tan\beta= 5$, $|\mu |=500$, $\text{  $|$}M_1\text{$|$ = } 130\text{ ,  $|$}M_2\text{$|$ = }110\text{ ,  }m_N\text{ = }260\text{ ,  }m_E\text{ = }240\text{ ,  }m_0\text{ = }7\times10^{4}\text{ ,  }m_0^{\tilde{\nu }}\text{ = }5\times10^4\text{ , $|$}A_0\text{$|$ = $|$}A_0^{\tilde{\nu }}\text{$|$ = }6\times 10^5\text{ , }\alpha _1\text{ = }0.4\text{ ,  }\alpha _2\text{ = }\alpha _{\mu }\text{ = }0.5\text{,  }\alpha _{A_0}\text{ = }\alpha _{A_0^{\tilde{\nu }}}\text{ = }1$.
The mixings are $\text{$|$}f_3\text{$|$ = }3\times 10^{-5}\text{ ,  $|$}f_3^{\prime}\text{$|$ = }4\times 10^{-6}\text{ ,  $|$}f_4\text{$|$ = $|$}f_4^{\prime}\text{$|$ = }0.8\text{ ,  $|$}f_4^{\prime\prime }\text{$|$ = }0.1\text{ ,  $|$}f_5\text{$|$ = }3\times10^{-6}\text{ ,  $|$}f_5^{\prime}\text{$|$ = }7\times10^{-6}\text{ ,  $|$}f_5{}^{\prime\prime }\text{$|$ = }5\times10^{-6}\text{ . }$Their CP phases are $\chi _3\text{ = }0.3\text{ ,  } $ $ \chi_3{}^{\prime}\text{ = }0.4\text{ ,  }$ $\chi _4\text{ = }1\text{ ,  }\chi _4^{\prime}\text{ = }\chi _4^{\prime\prime}\text{ = }0.5\text{ ,  }\chi _5\text{ = }\chi _5^{\prime}\text{ = }\chi _5{}^{\prime\prime}\text{ = }1$.
}

\label{fig5}
\end{figure}

\begin{table}\centering
\begin{tabular}{lll}

 & (i) $\chi_3^{\prime\prime} = 0.39$  & \hspace{0.4cm} (ii) $\chi_3^{\prime\prime} = 2.1$ \\  \hline
$F_{2\chi^{+}}^{\mu e}$ & $2.7\times10^{-18}\hspace{0.015cm}  e^{-0.74i}$ &\hspace{0.5cm}  $2.8\times10^{-18}\hspace{0.015cm}  e^{-2.1i}$  \\ 
$F_{2\chi^{0}}^{\mu e}$ & $3.4\times10^{-21}\hspace{0.015cm}  e^{0.56i}$ &\hspace{0.5cm}  $3.5\times10^{-21}\hspace{0.015cm}  e^{-0.39i}$  \\ 
$F_{2W}^{\mu e}$ & $5.8\times10^{-15}\hspace{0.015cm}  e^{-1.5i}$ &\hspace{0.5cm}  $6.1\times10^{-15}\hspace{0.015cm}  e^{0.12i}$  \\ 
$F_{2Z}^{\mu e}$ & $4.0\times10^{-15}\hspace{0.015cm}  e^{1.4i}$ &\hspace{0.5cm}  $4.7\times10^{-15}\hspace{0.015cm}  e^{-0.15i}$  \\ \hline

$F_2^{\mu e} (0)$ & $1.9\times10^{-15}\hspace{0.015cm}  e^{-1.1i}$ &\hspace{0.5cm}  $1.1\times10^{-14}\hspace{0.015cm}  e^{0.0024i}$  \\ 
${\cal{B}}_m$  & $5.3\times10^{-14}$ & \hspace{0.5cm}  $5.6\times10^{-13}$  \\ \hline\hline

$F_{3\chi^{+}}^{\mu e}$ & $2.4\times10^{-18}\hspace{0.015cm}  e^{3.04i}$ & \hspace{0.5cm}  $9.5\times10^{-19}\hspace{0.015cm}  e^{0.45i}$  \\ 
$F_{3\chi^{0}}^{\mu e}$ & $1.3\times10^{-20}\hspace{0.015cm}  e^{0.53i}$ & \hspace{0.5cm}  $1.6\times10^{-21}\hspace{0.015cm}  e^{-1.1i}$  \\ 
$F_{3W}^{\mu e}$ & $3.3\times10^{-15}\hspace{0.015cm}  e^{-1.7i}$ & \hspace{0.5cm}  $2.3\times10^{-15}\hspace{0.015cm}  e^{-0.019i}$  \\ 
$F_{3Z}^{\mu e}$ & $1.5\times10^{-15}\hspace{0.015cm}  e^{1.9i}$ & \hspace{0.5cm}  $5.8\times10^{-16}\hspace{0.015cm}  e^{0.34i}$  \\  \hline

$F_3^{\mu e} (0)$ & $2.2\times10^{-15}\hspace{0.015cm}  e^{-2.1i}$ & \hspace{0.5cm}  $2.9\times10^{-15}\hspace{0.015cm}  e^{0.051i}$  \\
${\cal{B}}_e$  & $6.6\times10^{-14}$ & \hspace{0.5cm}  $6.7\times10^{-14}$  \\ \hline\hline

\b & $1.2\times10^{-13}$ & \hspace{0.5cm} $1.7\times10^{-12}$  \\ 
\bb & $2.5\times10^{-19}$ & \hspace{0.5cm} $2.5\times10^{-19}$  \\ 
\hline\hline
$d_e$ ($e$cm) & $4.7\times10^{-29}$ & \hspace{0.5cm} $6.1\times10^{-30}$  \\ 
$m_{\nu3}$ & $4.7\times10^{-11}$ & \hspace{0.5cm} $4.7\times10^{-11} $ \\
$m_{\nu2}$ & $9.1\times10^{-12}$ & \hspace{0.5cm} $8.8\times10^{-12} $ \\
$m_{\nu1}$ & $ 2.1\times10^{-12}$ & \hspace{0.5cm} $4.9\times10^{-13} $ \\ 
%\hline
\hline\hline
\end{tabular}
\caption{An exhibition of the numerical values of the form factors $F_2^{\mu e}$  and $F_3^{\mu e}$ and their sub pieces for 
two cases: (i) and (ii). For case (i)  $|f_3^{\prime\prime}|=10^{-4}$ and $\chi_3^{\prime\prime} = 0.39$ while for case (ii)  $|f_3^{\prime\prime}|= 0.7\times10^{-4}$ and $\chi_3^{\prime\prime} = 2.1$. All other parameters used in this table are the same as Fig.\ref{fig5}.
The magnetic and electric transition operators ${\cal{B}}_m$  and ${\cal{B}}_e$ are also listed as are $d_e$ and 
and the  neutrino mass eigenstates for each case listed above. 
}
\label{table2}
\end{table}

 \begin{figure} \center
    \includegraphics[scale=.39]{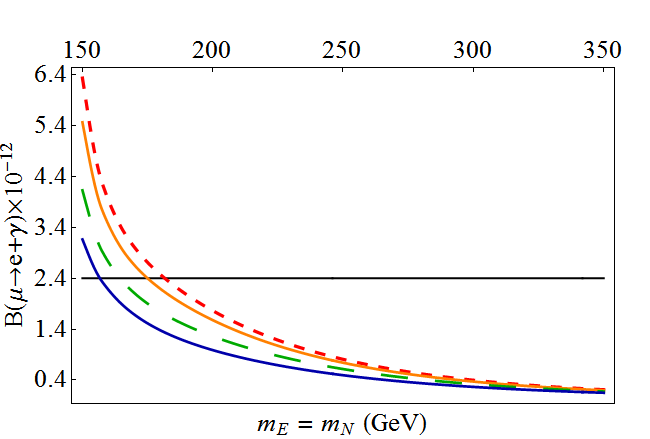}
       \caption{A plot of \b as a function of $m_E=m_N$ where the curves from top to bottom at $m_E=m_N=150$ are for values of $|f_4|=0.1$, $0.5$, $1$, $1.5$. The parameters used are $|\mu|=500$, $|M_1|=130$, $|M_2|=110$, $\tan\beta=10$, $m_0=4\times10^{4}$, $|A_0|= 6\times10^{5}$, $m_0^{\tilde{\nu }}=5\times10^4$, $|A_0^{\tilde{\nu }}|= 6\times10^{5}$, $\alpha _1=0.4$, $\alpha _2=0.5$, $\alpha _{\mu }=0.5$, $\alpha _{A_0}= \alpha _{A_0^{\tilde{\nu }}}=0.6$. 
The mixings are $|f_3|=5\times10^{-5}$, $|f_3^{\prime}|=5\times10^{-5}$, $|f_3^{\prime\prime}|=5\times10^{-5}$, $|f_4'|=0.5$, $|f_4''|=5\times10^{-1}$, $|f_5|=3\times10^{-6}$, $|f_5'|=8\times10^{-7}$, $|f_5''|=5\times10^{-6}$, $\chi_3=\chi_3'=\chi_3''=0.3$, $\chi_4=\chi_4'=\chi_4''=1$, $\chi_5=\chi_5'=\chi_5''=1$.
}

\label{fig6}
\end{figure}

\end{document}